\documentclass[a4paper,fleqn,usenatbib]{mnras}

\usepackage[T1]{fontenc}
\usepackage{ae,aecompl}

\usepackage{graphicx}	% Including figure files
\usepackage{amsmath}	% Advanced maths commands
\usepackage{amssymb}	% Extra maths symbols

\title[An extinction free AGN selection by 18-band SED fitting in mid-infrared in the AKARI NEP deep field]{An extinction free AGN selection by 18-band SED fitting in mid-infrared in the AKARI NEP deep field}

\author[Ting-Chi Huang]{Ting-Chi Huang,$^{1}$\thanks{E-mail: s104022505@m104.nthu.edu.tw (NTHU)}
Tomotsugu Goto,$^{1,2}$ Tetsuya Hashimoto,$^{1,2}$ Nagisa Oi$^{3,4}$
\newauthor and Hideo Matsuhara$^4$ 
\\
\\
$^{1}$Department of Physics, National Tsing Hua University, No. 101, Section 2, Kuang-Fu Road, Hsinchu City 30013, Taiwan\\
$^{2}$Institute of Astronomy, National Tsing Hua University, No. 101, Section 2, Kuang-Fu Road, Hsinchu City 30013, Taiwan\\
$^{3}$Tokyo University of Science, 1-3 Kagurazaka, Shinjuku-ku, Tokyo 162-8601, Japan\\
$^{4}$Department of Infrared Astrophysics, Institute of Space and Astronautical Science, Japan Aerospace Exploration Agency (JAXA), \\
3-1-1 Yoshinodai, Chuo-ku, Sagamihara 252-5210 Japan
}

\date{Accepted XXX. Received YYY; in original form ZZZ}

\pubyear{2017}

\begin{document}
\label{firstpage}
\pagerange{\pageref{firstpage}--\pageref{lastpage}}
\maketitle

\begin{abstract}
We have developed an efficient Active Galactic Nucleus (AGN) selection method using 18-band Spectral Energy Distribution (SED) fitting in mid-infrared (mid-IR). AGNs are often obscured by gas and dust, and those obscured AGNs tend to be missed in optical, UV and soft X-ray observations. Mid-IR light can help us to recover them in an obscuration free way using their thermal emission. On the other hand, Star-Forming Galaxies (SFG) also have strong PAH emission features in mid-IR. Hence, establishing an accurate method to separate populations of AGN and SFG is important. However, in previous mid-IR surveys, only 3 or 4 filters were available, and thus the selection was limited. We combined AKARI's continuous 9 mid-IR bands with WISE and Spitzer data to create 18 mid-IR bands for AGN selection. Among 4682 galaxies in the AKARI NEP deep field, 1388 are selected to be AGN hosts, which implies an AGN fraction of 29.6$\pm$0.8$\%$ (among them 47$\%$ are Seyfert 1.8 and 2). Comparing the result from SED fitting into WISE and Spitzer colour-colour diagram reveals that Seyferts are often missed by previous studies. Our result has been tested by stacking median magnitude for each sample. Using X-ray data from Chandra, we compared the result of our SED fitting with WISE's colour box selection. We recovered more X-ray detected AGN than previous methods by 20$\%$.

\end{abstract}

\begin{keywords}
Galaxies: Infrared -- Galaxies: Active -- Galaxies: Star Formation
\end{keywords}

%%%%%%%%%%%%%%%%%%%%%%%%%%%%%%%%%%%%%%%%%%%%%%%%%%

%%%%%%%%%%%%%%%%% BODY OF PAPER %%%%%%%%%%%%%%%%%%

\section{Introduction}
Active galactic nucleus (AGN) is important in galaxy evolution from many aspects. It has been widely believed that there is a supermassive black hole (SMBH) in every AGN, and the black hole mass are found to be related to the bulge mass \citep[e.g.,][]{Magorrian et al. 1998}. AGN is in a phase that a large amount of materials are accreting onto the SMBH in the centre of galaxy, and huge amount of energy is released from it. Hence, by investigating accretions onto SMBHs and energy input to galaxies, we are able to understand how galaxies and SMBHs have evolved and grown with cosmic time. Additionally, it has been proposed that high redshift AGN contributes to the cosmic reionization \citep[e.g.,][]{Glikman et al. 2010,Madau P. and Haardt F. 2015}. As a consequence, it is essential to have a method to select a complete sample of AGN in a consistent manner. 

However, because of the obscuration of dust and gas, there are many AGNs missed in optical, UV or soft X-ray observations \citep[e.g.,][]{Alexander et al. 2001,Richard et al. 2003,Webster et al. 1995}. Fortunately, those obscured AGN can be probed by infrared (IR) due to its warm dust emission. But not only AGN, star-forming galaxies (SFGs) are also prevalent in IR observations, so separating the populations of these two sources is extremely important in IR survey. In mid-IR, SFG has polycyclic aromatic hydrocarbon (PAH) emission features at 3.3, 6.2, 7.7, 8.6, and 11.3 $\mu$m. On the other hand, AGN has a power law spectrum from black radiation in mid-IR, so we can select AGN by mid-IR colours \citep{Jarrett et al. 2011,Lacy et al. 2004,Richard et al. 2006,Stern et al. 2005}. However, those previous studies selected AGN by mid-IR colours from observations of  \textit{Wide field Infrared Survey Explorer} (WISE) and \textit{Spitzer infrared telescope} (Spitzer) with only 3 or 4 bands of flux/magnitude. Not only those previous work are limited by available filters, but there is also a gap of wavelength range between mid-IR filters. For example,  Spitzer IRAC has 4 bands with effective wavelength at 3.6, 4.5, 5.8 and 8.0 $\mu$m. However, the next longer band they have is Spitzer MIPS24, of which effective wavelength is at 24~$\mu$m. They lack the band from 8 to 24 $\mu$m, which is an important range for the diagnosis of AGN and SFG at redshift from 0.5 to 1.5. 

Unlike previous mid-IR AGN selection methods using colour-colour diagram of Spitzer and WISE, we used AKARI, which is a Japanese IR space telescope with 9 mid-IR filters, to select AGN \citep{Murakami et al. 2007}. AKARI has 9 continous bands which cover the whole mid-IR range from 2 to 24 $\mu$m \citep{Matsuhara et al. 2006}. Having these 9 continous bands is an advantage of AKARI, and by utilizing this, we perform SED fitting to select AGN. For example, \citet{Chung et al. 2014} shows that SED fitting can successfully select AGN populations, but they have only 5 bands in mid-IR with 3 galaxy and 1 AGN templates for fitting. In this work, we combine AKARI, WISE and Spitzer, having total 18 bands in mid-IR, which reveals the features of AGN and SFG more accurately, and allows us to fit with 25 empirical models.  

This paper is organised as follows. We describe our data, analysis, and sample construction in Sect.~\ref{sec:DandA}. Our results of AGN selection and examinations are described in Sect.~\ref{results}. In Sect.~\ref{discussion}, we present uncertainties of our method, comparisons with other work, conculsions and future prospects. This paper ends up with a summary in Sect.~\ref{summary}.

\section{Data and Analysis}  
\label{sec:DandA}
\subsection{Data}
\label{sec:datas} % used for referring to this section from elsewhere

In this research, we used the catalogue from the North Ecliptic Pole (NEP) deep field survey of AKARI infrared astronomical satellite \citep{Takagi et al. 2012,Murata et al. 2013}. The NEP deep survey observed 0.57 deg$^2$ area at (R.A. = 17$^h$56$^m$, Dec. = 66$^\circ$37') and was carried out by AKARI Infrared Camera (IRC) which has 3 channels, NIR, MIR-S and MIR-L. There are 3 filters in each channel, so the AKARI IRC has 9 filters in total: N2, N3, N4, S7, S9W, S11, L15, L18W, and L24. The filters correspond to 2.4, 3.2, 4.1, 7, 9, 11, 15, 18, and 24~$\mu$m of the reference wavelengths. The 5$\sigma$ detection limits are 11, 9, 10, 30, 34, 57, 87, 93, and 256 $\mu$Jy in the above 9 bands respectively \citep{Murata et al. 2013}. For the exact detection criteria, please refer \citet{Murata et al. 2013}. The advantage of AKARI IRC is that it has continous coverage from near to mid-infrared (2 to 24~$\mu$m). These bands can capture AGN's warm dust radiation and the PAH emissions, one of which is the feature of SFG at 7.7~$\mu$m (at redshift 0.5 to 1.5). There are 27770 objects in the full catalogue, but in order to have information in MIR, we required the objects to have L18W band flux detection, and 5761 objects remained. Most of them are regarded as point sources since the size of point spread function is 6 arcsec in 18 $\mu$m, which corresponds to 27 kpc at z = 0.3.

Among them, $12\%$ and $25\%$ objects have UV observation by GALEX FUV and NUV band, respectively \citep{Buat et al. 2015}. More than $80\%$ objects have optical photometries in $u^*$, $g'$, $r'$, $i'$, and $z'$-bands from CFHT MegaCam, and $Y$, $J$ and $K_{s}$-bands from CFHT WIRCam \citep{Oi et al. 2014}. These UV and optical information help us fit the stellar component in galaxy spectra. For mid-IR range, there are $45\%$ objects with WISE observation, and $6\%$ with Spitzer \citep{Jarrett et al. 2011}. Combining the magnitudes from AKARI's 9 bands with WISE's 4 bands (3.4, 4.6, 8, and 22 $\mu$m), Spitzer IRAC's 4 bands (3.6, 4.5, 5.8, and 8.0 $\mu$m), and one band of Spitzer MIPS24 (24$\mu$m), we have at most 18-band magnitudes in mid-IR, which are used in our AGN selection. Also, $3\%$ and $5\%$ objects are observed by Herschel PACS and SPIRE, respectively \citep{Buat et al. 2015}. 

We converted all the magnitudes into AB system and cross-matched them by coordinate with 1 arcsec tolerance radius. We tried to match as many AKARI objects as possible with the ALLWISE and other external catalogues. AKARI, WISE, Spitzer and Herschel used different aperture radii for photometry. Only a small number of objects have Spitzer or Herschel data, so the effect of different aperture radius from them is small due to its sample size. For AKARI and WISE, the aperture radii of AKARI's NIR and MIR bands are 6.3 and 6.0 arcsec \citep{Murata et al. 2013}, and WISE's first three bands and the fourth band use 8.25 and 16.5 arcsec \citep{Wright et al. 2010}. The flux deviation at 24$\mu$m, for example, between AKARI L24 and WISE [24] is estimated to be $\sim$20$\%$ ($\sim$0.19 magnitude), however, which is smaller than the observational errors ($\sim$0.32 magnitude in average).  Therefore, we believe that the difference in aperture size does not cause large uncertainties in this study.  

\begin{figure}
	\includegraphics[width=\columnwidth]{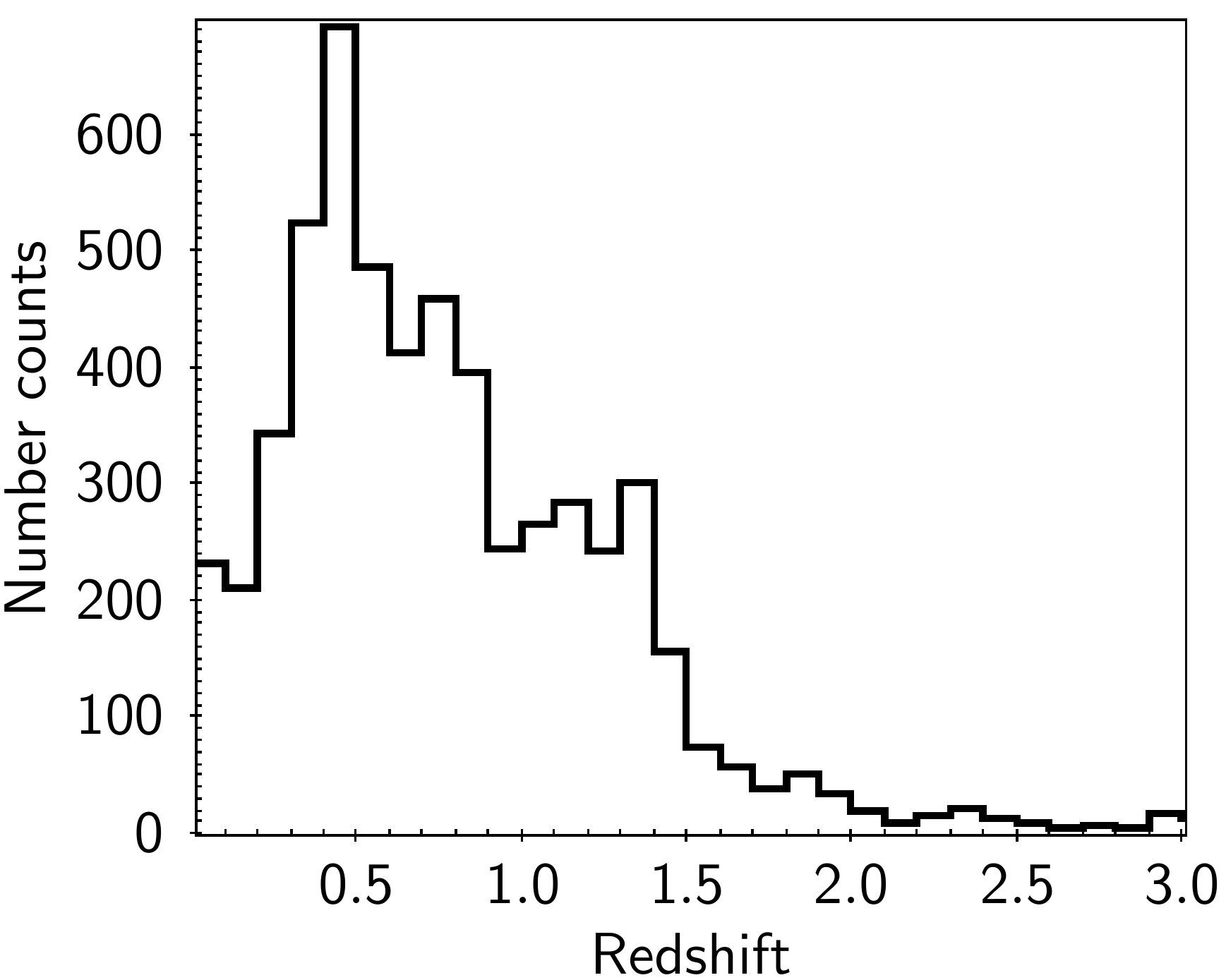}
    \caption{The photometric redshift distribution of the L18W detected sample in the AKARI NEP deep field. The bin size in this figure is 0.1.}
    \label{fig:redshift}
\end{figure}

The photometric redshifts of AKARI NEP deep sources are constructed and described in \citet{Oi et al. 2014}. They computed photometric redshifts by the $LePhare$ Code \citep{Arnouts et al. 1999} with 62 galaxy templates and 154 star templates. In our sample, the photometric redshift distribution is shown in Fig.~\ref{fig:redshift}. The redshift of the sample is ranging from 0 to 6, but there are few objects at high redshifts. Hence, we only plot the redshift distribution from 0 to 3 and the objects mainly distribute from 0 to 1.5. Besides, there are 457 AGNs with X-ray observation from Chandra which are included as the AGN benchmark to examine our selecting method \citep{Krumpe et al. 2015}. 

\subsection{SED fitting}
\label{SED}
Our main goal is to separate AGN and SFG in large galaxy populations from IR photometric data. We can achieve this by SED fitting because AGN and SFG have different features in mid-infrared. SFGs have strong PAH emission features at 6.2, 7.7, 8.6 and 11.3 $\mu$m, while AGN shows continuous strong continuum due to black body radiation from warm dust (T$\sim$1000K). Our SED fitting was performed using $LePhare$ Code\footnote{\url{http://www.cfht.hawaii.edu/~arnouts/LEPHARE/lephare.html}} \citep{Arnouts et al. 1999}. We used the Spitzer Wide-area Infrared Extragalatic (SWIRE) templates from \citet{Polletta et al. 2007}, which includes 16 galaxy and starburst templates and 9 AGN templates (Table~\ref{tab:model}). We sorted the models with increasing integers from SFG to AGN. In other words, the model with number from 1 to 16 are the SEDs of SFG, and the SEDs of AGN are represented by the models with the number from 17 to 25. A correction factor was applied to the filter calibration in $LePhare$ with the parameter FILTER$\_$CALIB=1, which assumes the product of frequency and reference spectrum is a constant and is suitable for calibrating mid-infrared filters. The extinction law from \citet{Calzetti et al. 2000} has been included with colour excess parameter EB$\_$V=0 to 3. Our fitting applied two libraries, which are a galaxy library and a far infrared (FIR) library. The galaxy library described the stellar component of galaxy spectrum \citep[][CWW-Kinney throughout this paper]{Coleman et al. 1980,Kinney et al. 1996}. The FIR library was used to select AGN \citep{Polletta et al. 2007}. In the FIR analysis with $LePhare$, the stellar component was substracted according to the best-fitting model by setting parameter FIR$\_$SUBSTELLAR=YES, and then the remaining spectrum is constructed from SF and AGN. For this, we used 3~$\mu$m as a lower rest frame wavelength constraint in the FIR SED fitting.  Since the photometric redshift had been already measured in our sample \citep{Oi et al. 2014}, we set the parameter ZFIX=YES. Then the code would use the premeasured redshift, fit the SED and search the best model. Finally, we separated AGN and SFG by checking the best-fitting model of each object.

After the SED fitting, we applied some basic criteria to select the sample so that we could eliminate the objects without enough information. (1) An object must have a best-fitting model. (2) An object must have a redshift value. (3) The number of bands used in FIR SED fitting must equal or larger than 3.  With the above criteria, we obtained the catalogue with 4833 galaxies, and we call it "the SED sample" throughout this paper. In order to compare the AGN selection by SED fitting with previous selections by IR colour-colour diagram, we also constructed "the colour sample" by the objects also detected with WISE. There are total 2387 objects in this colour sample. All these objects are detected in all 4 bands of WISE.

\begin{table}
	\centering
	\caption{Templates from \citet{Polletta et al. 2007} that we use as a far infrared SED library. We regard the galaxy as AGN if the best-fitting template is 17 or above. }
	\label{tab:model}
	\begin{tabular}{lccr} % four columns, alignment for each
		\hline
		Model & Spectral Type \\
		\hline
		1 & Elliptical (t=13 Gyr)  \\
		2 & Elliptical (t= 2 Gyr) \\
		3 & Elliptical (t= 5 Gyr) \\
		4 & Spiral 0 \\
		5 & Spiral a \\
		6 & Spiral b \\
		7 & Spiral c \\
		8 & Spiral d \\
		9 & Spiral dm \\
		10 & Spi4 (Spiral c) \\
		11 & Arp220 (Starburst/ULIRG) \\
		12 & IRAS 20551-4250 (Starburst/ULIRG) \\
		13 & IRAS 22491-1808 (Starburst/ULIRG) \\
		14 & M82 (Starburst) \\
		15 & NGC 6240 (Starburst/Sey2) \\
		16 & NGC 6090 (Starburst) \\
		17 & Seyfert 1.8 \\
		18 & Seyfert 2  \\
		19 & QSO2 (Type-2 QSO) \\
		20 & QSO1 (Type-1 QSO) \\
		21 & BQSO1 (Type-1 QSO) \\
		22 & TQSO1 (Type-1 QSO) \\
		23 & Mrk231 (Seyfert 1, BAL QSO, Starburst/ULIRG)  \\
		24 & IRAS19254-7245 South (Seyfert 2+Starburst/ULIRG) \\
		25 & Torus (Type 2 QSO)\\
		\hline
	\end{tabular}
\end{table}

\section{Results}
\label{results}
\subsection{AGN Selection}
\label{AGN}
\subsubsection{SED AGN}
\label{SED AGN}
\begin{figure}
	\includegraphics[width=\columnwidth]{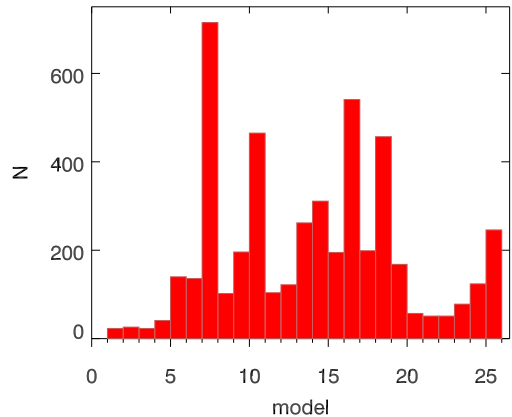}
    \caption{The distribution of the best-fitting models in the SED sample. The corresponding model numbers are shown in Table~\ref{tab:model}. AGN models are number 17 to 25. We regard an object as AGN if its model number is larger than 16.}
    \label{fig:mod}
\end{figure}

From the SED sample we constructed in section~\ref{SED}, we first separated stars and galaxies by their CLASS$\_$STAR parameter from SExtractor in $z$-band image of CFHT MegaCam. \cite{Oi et al. 2014} suggests that it is feasible to separate the stars and galaxies with (1) the CFHT $u^*-J$ versus $g'-K_s$ colour-colour diagram; (2) the value > 0.8 of CLASS$\_$STAR parameter from SExtractor; and (3) the CFHT colour $u^*-g'$ > 0.4. In this paper, we simply use the value > 0.9 of CLASS$\_$STAR parameter as our star-galaxy separation, because this selection is consistent with above (1) and (2) combined, which is the result of \cite{Oi et al. 2014}. Without the criterion (3), some quasars could be misclassified as stars. However, there are only a small number of stars in our data, so we do not think this approximation would change our result too much. After the separation, we then selected AGN in the galaxy sample by checking the best-fitting model. The distribution of the best-fitting models of all objects is shown in Fig.~\ref{fig:mod}. There are 4682 galaxies in total, and 1388 AGNs are selected, that is, the AGN fraction is 29.6$\pm$0.8$\%$. In some cases the SEDs of objects are close to multiple templates, that is, the $\chi^2$ values do not differ so much from the best-fitting model to some other models for some objects. Nevertheless, we show that this separating method is still statistically valid by examinations in next sections.   

\subsubsection{Colour AGN}
\label{Colour AGN}    
\begin{figure}
	\includegraphics[width=\columnwidth]{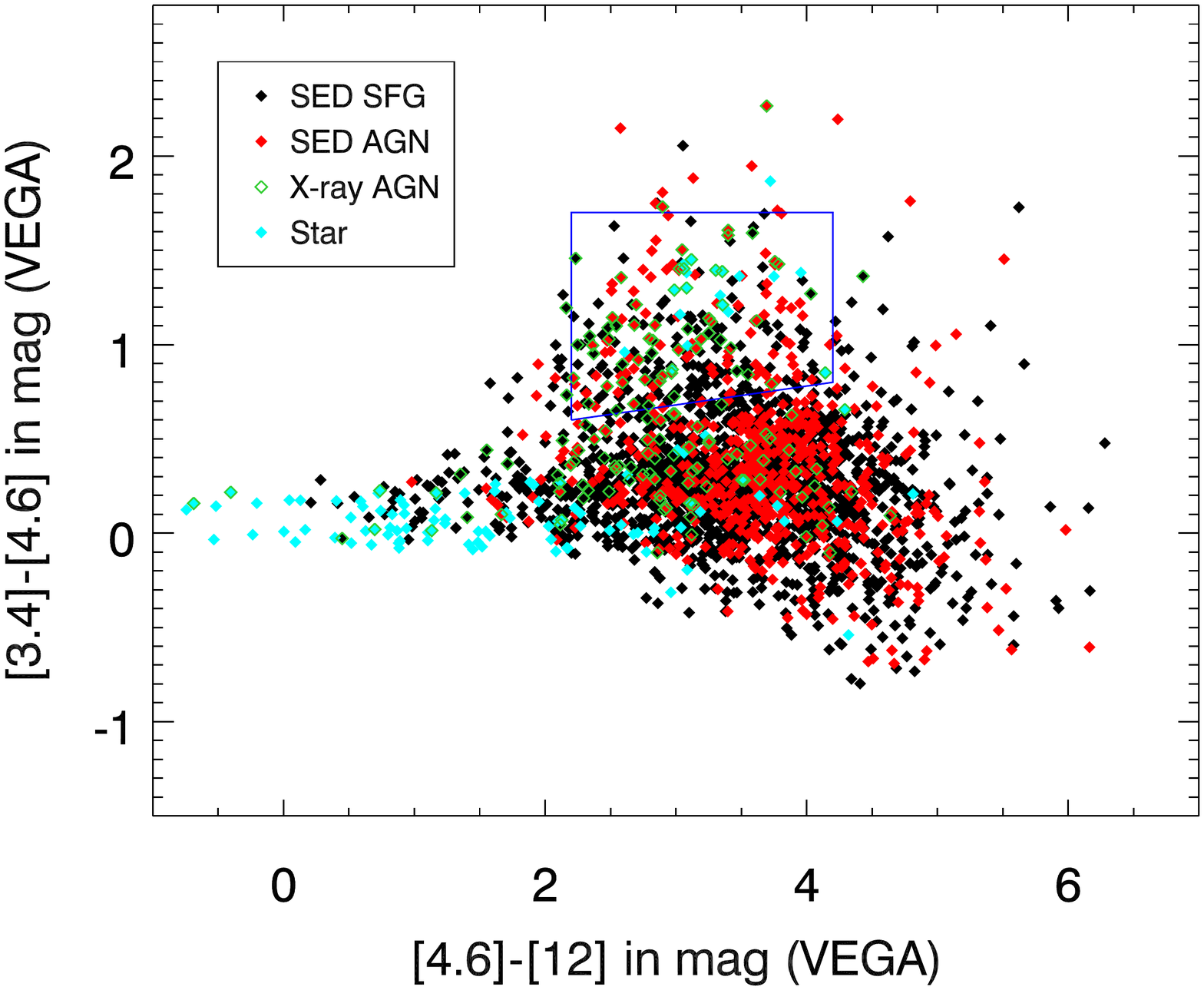}	
	\includegraphics[width=\columnwidth]{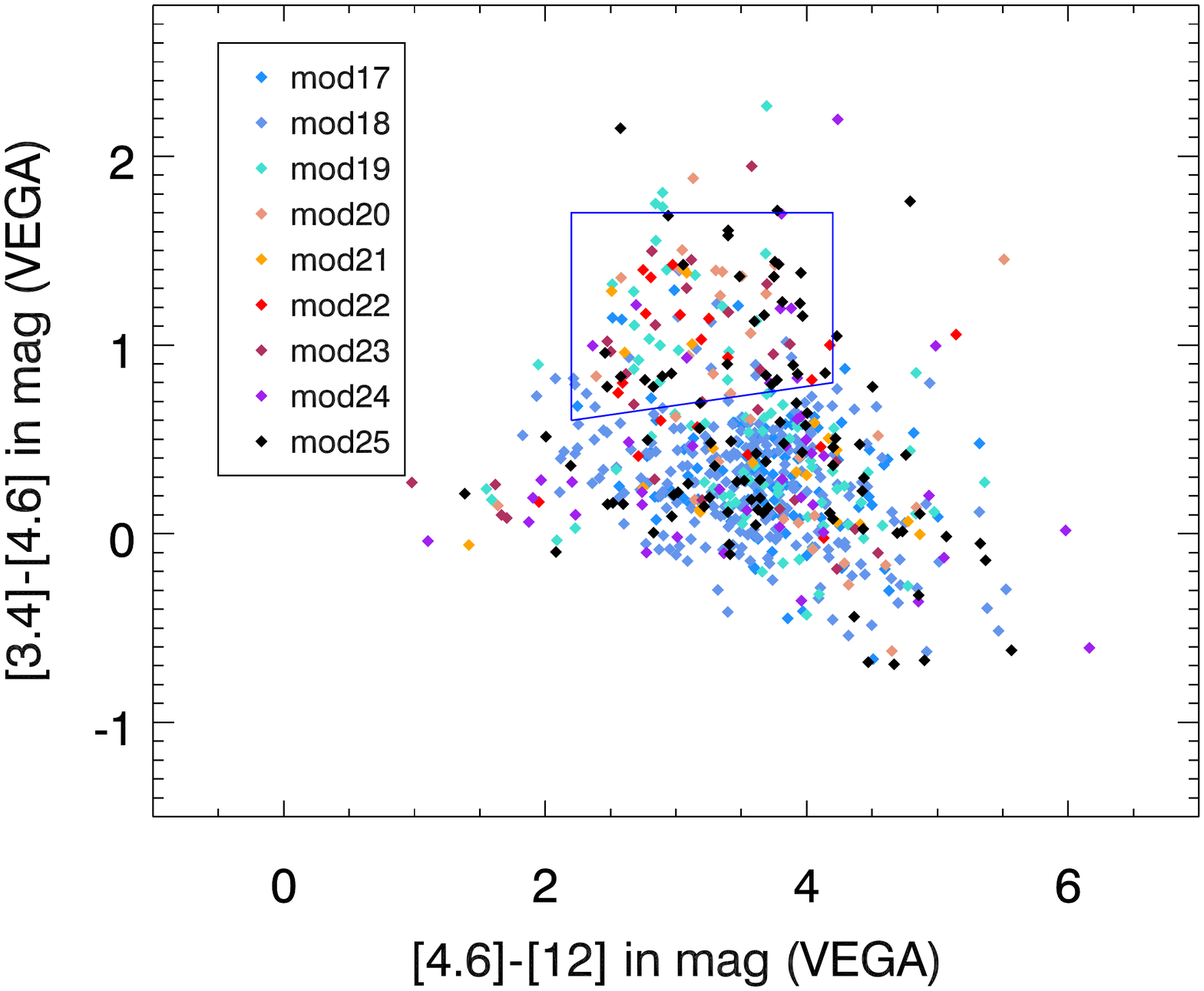}
    \caption{The WISE colour-colour diagram in VEGA magnitude system. \textsl{Top:~}The black diamonds represent galaxies and the red diamonds are AGNs selected by SED fitting in 18 bands with the template from \citet{Polletta et al. 2007}. Green empty diamonds are AGNs confirmed by Chandra observation in X-ray. The cyan dots are stars, chosen by their CLASS$\_$STAR parameter ( > 0.9). The blue-lined box is the AGN criteria from \citet{Jarrett et al. 2011} \textsl{bottom:~}The WISE colour-colour diagram of SED AGNs with our best-fitting models. Different colour which was used for each model is shown in the legend (17 to 25 as in Table~\ref{tab:model})}
    \label{fig:ccdiagram}
\end{figure}

\begin{figure}
	\includegraphics[width=\columnwidth]{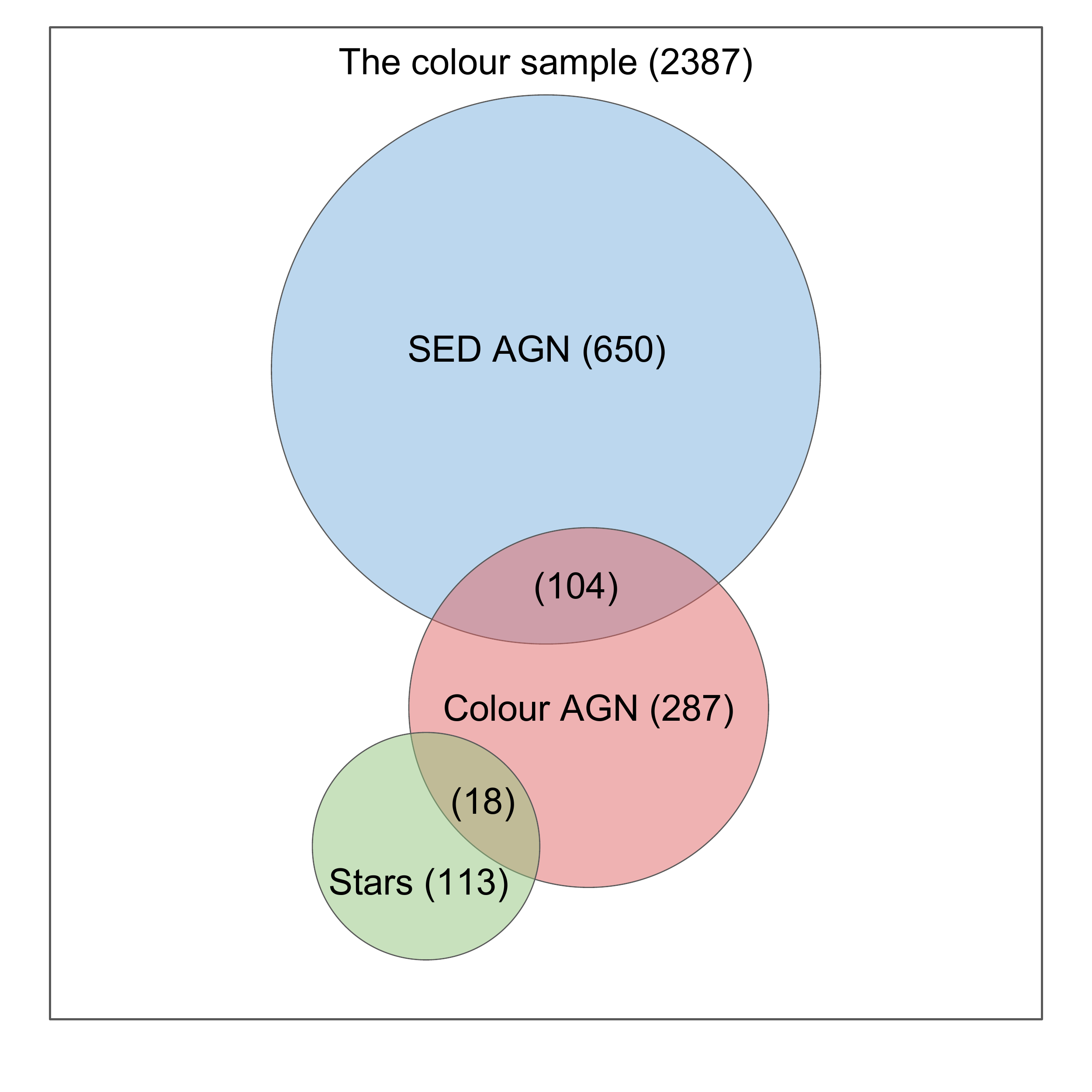}
    \caption{An illustration of Venn diagram for Table~\ref{tab:18-band}. The number of objects in each subset is written in parentheses. The areas of circles are drawn to scale, but overlapping areas are not.}
    \label{fig:Venn}
\end{figure}

The colour sample was used to compare our selection with previous IR selections in the colour-colour diagram of WISE bands, which is shown in the top panel of Fig.~\ref{fig:ccdiagram} in VEGA magnitude system \citep{Jarrett et al. 2011}. The colour selection box is empirically defined by the covering regions for different galaxy types. The different galaxy types are defined by the template from SWIRE \citep{Polletta et al. 2006,Polletta et al. 2007}. Their box is defined to cover QSOs, so inevitably they miss ULIRGs and host-dominated AGNs in redder and bluer part of the box. The bottom panel shows how the best fitting models distribute in the colour-colour diagram. Not surprisingly, most Seyfert galaxies (model 17 and 18) reside out of the box. The objects inside the box are mainly QSOs (model 19, 20, 21 and 22). There are 2387 objects with WISE magnitudes in the subsample that we have constructed. Among them, 113 objects have been classified to be stars by their CLASS$\_$STAR parameter. From now on, to avoid the confusion, we call the AGN and SFG selected by the SED fitting as "SED AGN" and "SED SFG", while the AGN and SFG selected by the WISE colour criteria are called as "colour AGN" and "colour SFG". By SED fitting, we selected 650 SED AGNs and 1624 SED SFGs. On the other hand, 287 colour AGNs are selected by the colour criteria and 104 objects satisfied the AGN selections of both method. We summarize this statistics in Table~\ref{tab:18-band} and illustrate in Fig.~\ref{fig:Venn}.

\begin{figure}
	\includegraphics[width=\columnwidth]{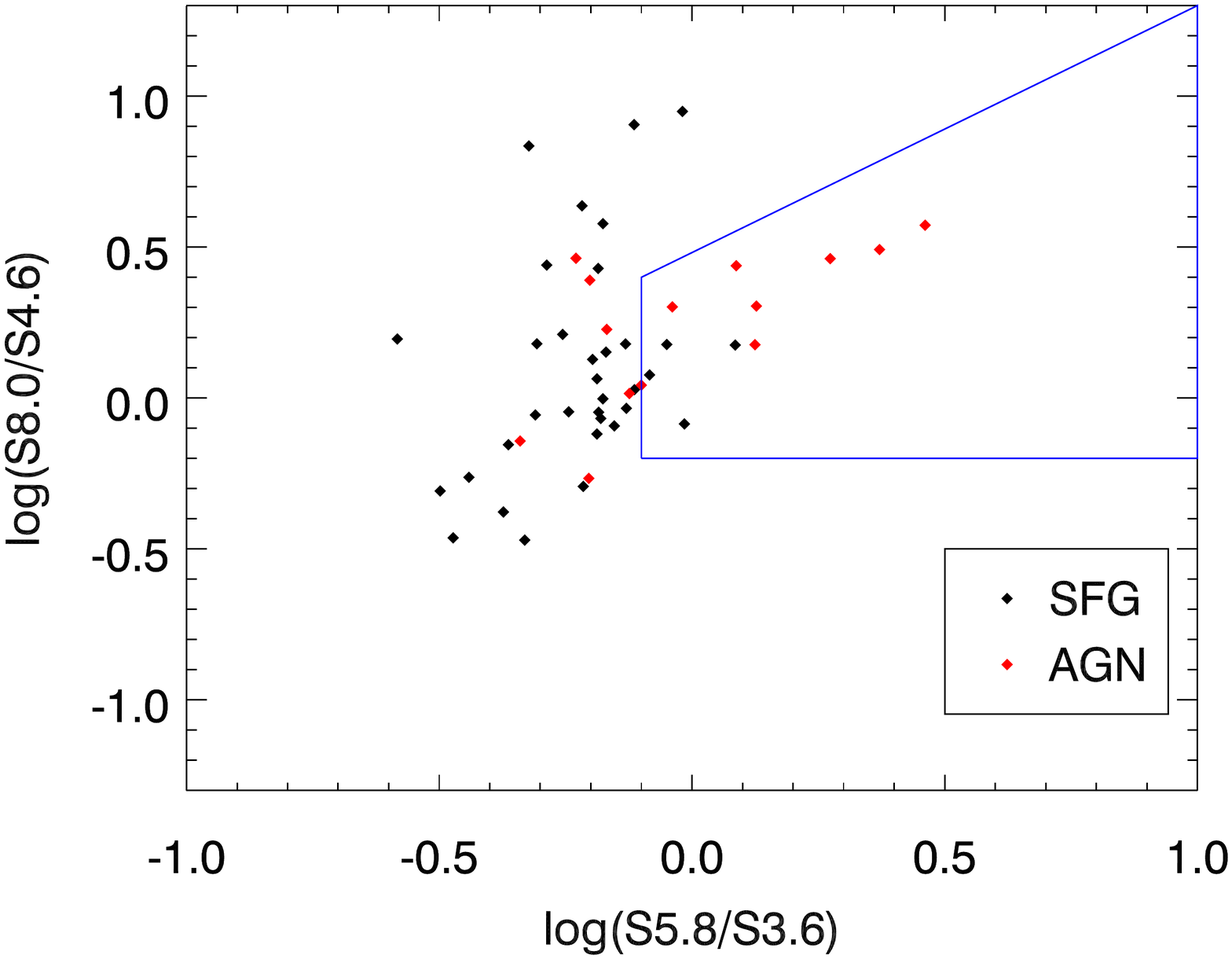}	
	\includegraphics[width=\columnwidth]{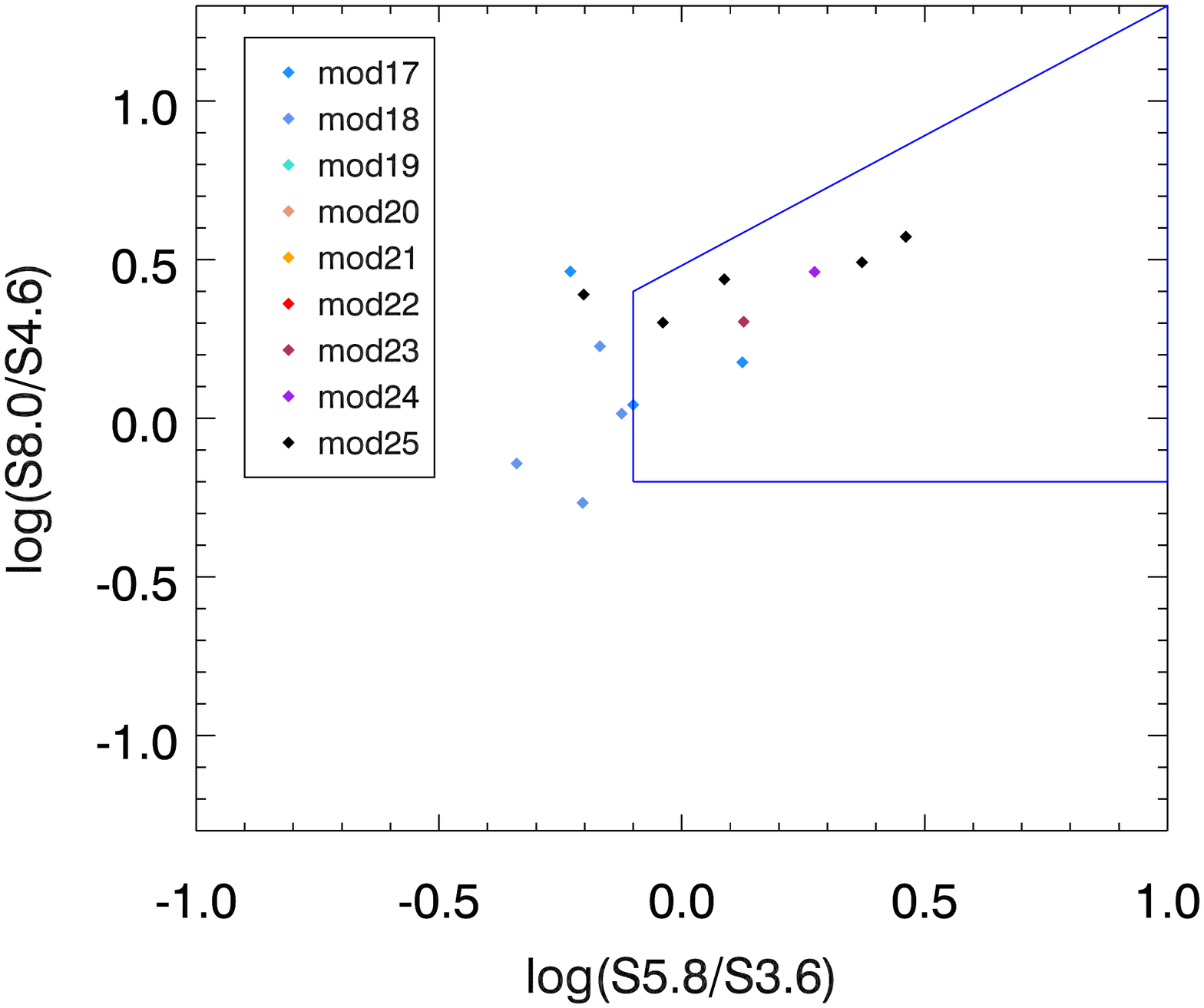}
    \caption{The Spitzer AGN selection diagrams from \citet{Lacy et al. 2007}. \textsl{Top:~}The black diamonds represent galaxies and the red diamonds are AGNs selected by SED fitting in 18 bands with the template from \citet{Polletta et al. 2007}. The blue-lined box is the AGN criteria. \textsl{bottom:~}The Spitzer AGN selection diagram of SED AGNs with our best-fitting models. Every model (17 to 25 as in Table~\ref{tab:model}) is in different colour as shown in the legend.}        
    \label{fig:ccdiagram_lacy}
\end{figure}

In addition to WISE, we also compare the AGN selection result with studies from Spitzer. However, we do not have as many counterparts of Spitzer in our catalogue, so the result is limited by the sample size. In total we have about 300 objects observed by Spitzer, but only 48 have been observed in Spitzer IRAC all 4 bands. In this small sample, 14 AGNs have been selected. The result is compared with the AGN selection box from \citet{Lacy et al. 2007} in Fig.~\ref{fig:ccdiagram_lacy}. The top panel shows the comparison of AGN selection, and each model is plotted separately in the bottom panel. Again, we confirm that our SED fitting selects Seyferts which are missed by the colour box.

\begin{table}
	\centering
	\caption{The statistics of AGN selections by SED fitting and IR colour criteria in the colour sample.}
	\label{tab:18-band}
	\begin{tabular}{lcccr} % five columns, alignment for each
		\hline
		 & SED SFG & SED AGN & Star & Total\\
		\hline
		colour SFG & 1459 & 546 & 95 & 2100\\
		colour AGN & 165 & 104 & 18 & 287\\
		Total & 1624 & 650 & 113 & 2387\\
		\hline
	\end{tabular}
\end{table}
    
\subsection{Examination by X-ray AGN}
We cross-correlated the X-ray catalogue from Chandra NEP deep survey~\citep{Krumpe et al. 2015} with our SED sample to check our selected AGN. The Chandra observation was performed from December 2010 to April 2011. They performed 15 ACIS-I pointings with total exposure time of 302 ks. About 0.34 deg$^2$ area was observed around (R.A. = 17$^h$55$^m$24$^s$, Dec. = 66$^\circ$33'33"). The full X-ray catalogue has 457 objects and there are 254 objects matched with the SED sample. We regard all these 254 objects as AGN. Hereafter, we call them the X-ray AGN sample. The main idea of this examination is to compare the results of AGN selection between SED fitting and IR colour criteria in the X-ray AGN sample by examing how many AGNs each method can recover. One may doubt the X-ray AGN assumption, because it is not necessary that every X-ray object is AGN and emits strongly in infrared. We test this reliability in the next section by stacking SEDs.  

\begin{figure}
	\includegraphics[width=\columnwidth]{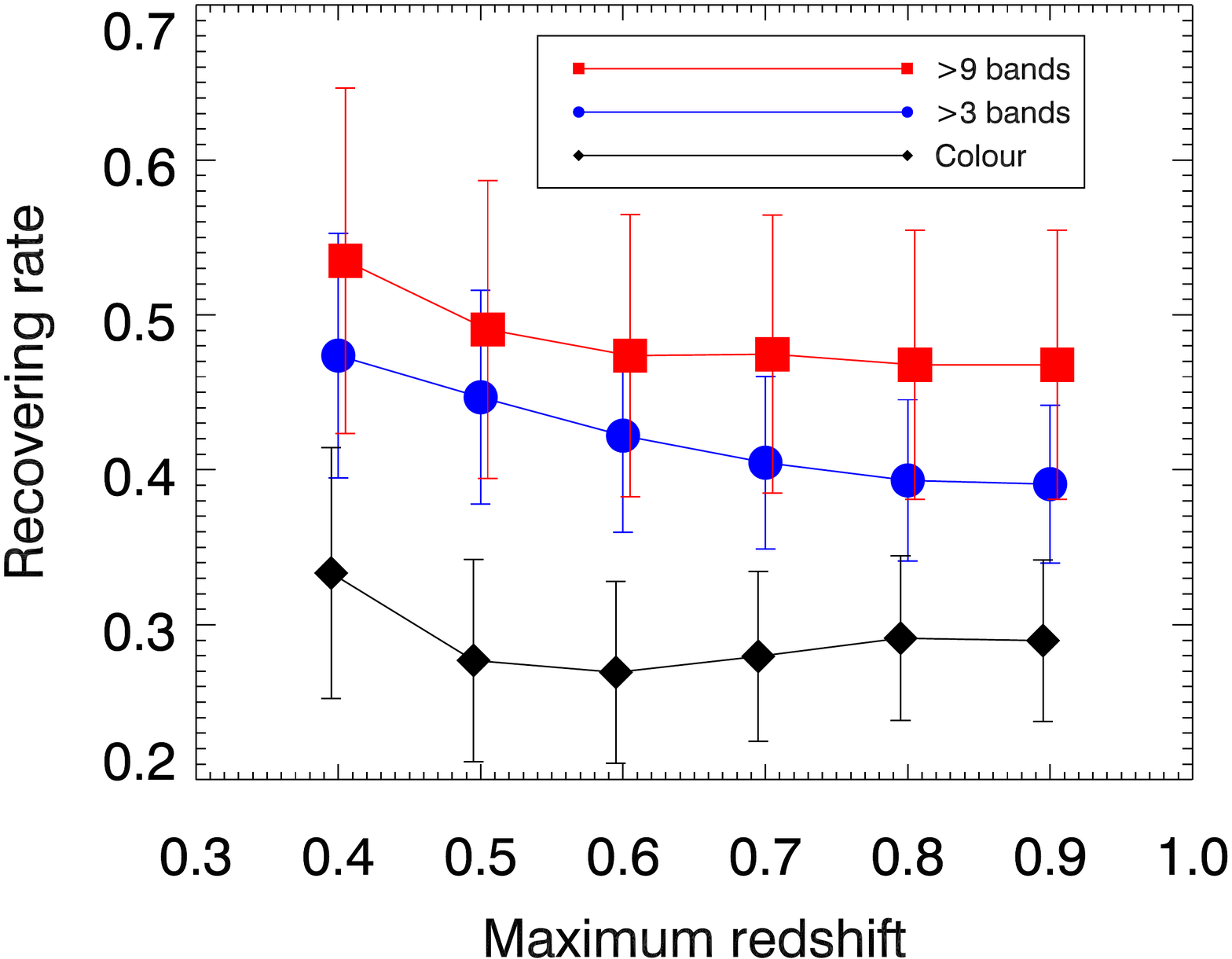}
	    \caption{The X-ray AGN recovering rate as functions of redshift. Results from colour selection is plotted in black and results from SED fitting are plotted in blue and red.}
    \label{fig:recovering_rate}
\end{figure}

The X-ray AGN recovering rate is higher in our SED fitting method as shown in Fig.~\ref{fig:recovering_rate}. Red squares and blue circles are the results by SED fitting, while the result by colour selection is plotted in black diamond. We separated the X-ray AGN sample with redshift and the number of bands used in fitting. The recovering rate slightly decreases with increasing redshift. Namely, it means that the recovering rate is higher in lower redshift samples. This can be explained by the inaccuracy of photometric redshift at high redshift and fewer bands in mid-IR at high redshift. As expected, the sample that has more than 9 bands in fitting recovers more X-ray AGN than those with only 3 bands. In either case, SED fitting recovers more X-ray AGN than the colour-colour selection. 

\subsection{Examination by stacked median magnitude}
\begin{figure}
	\includegraphics[width=\columnwidth]{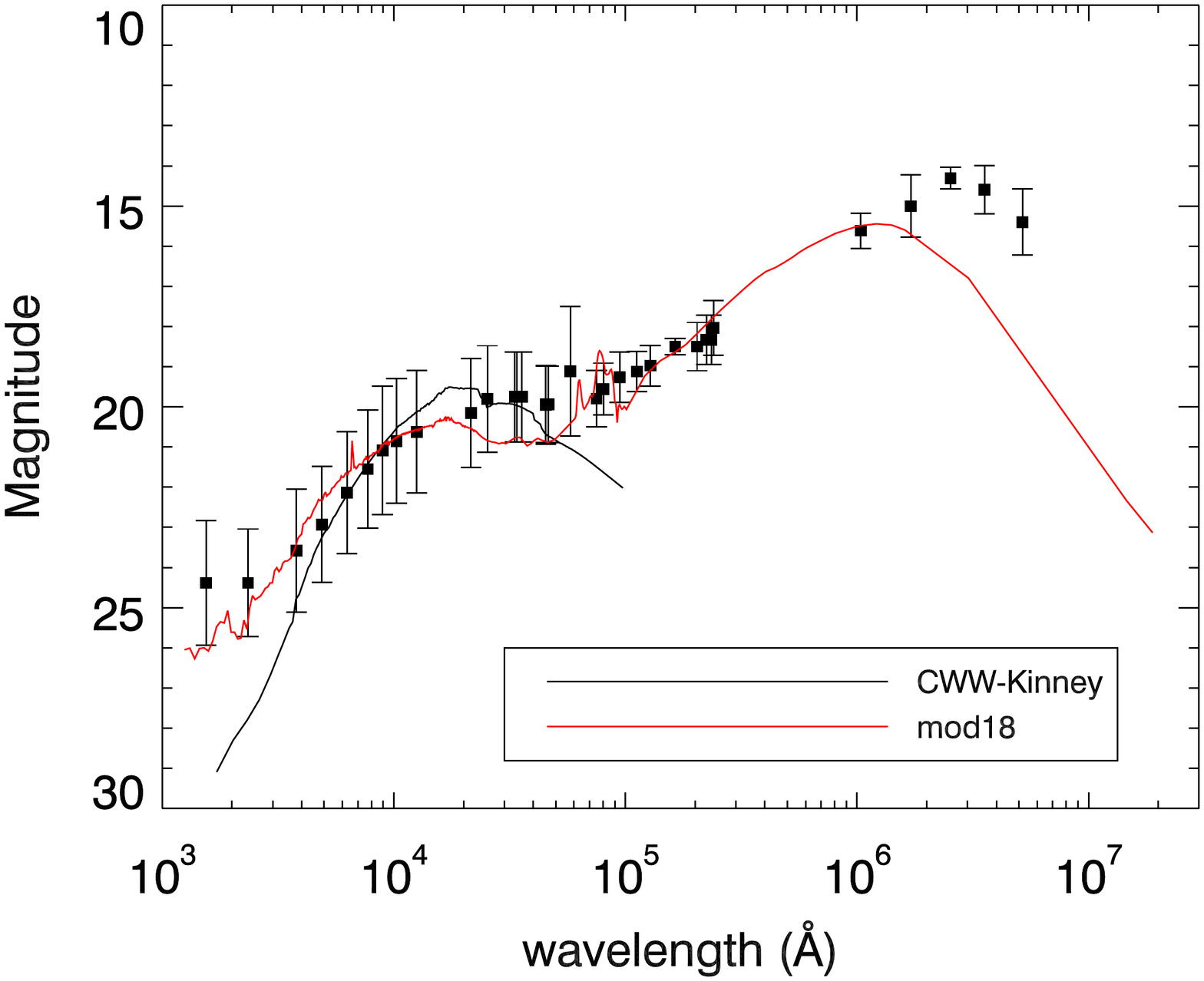}
	\includegraphics[width=\columnwidth]{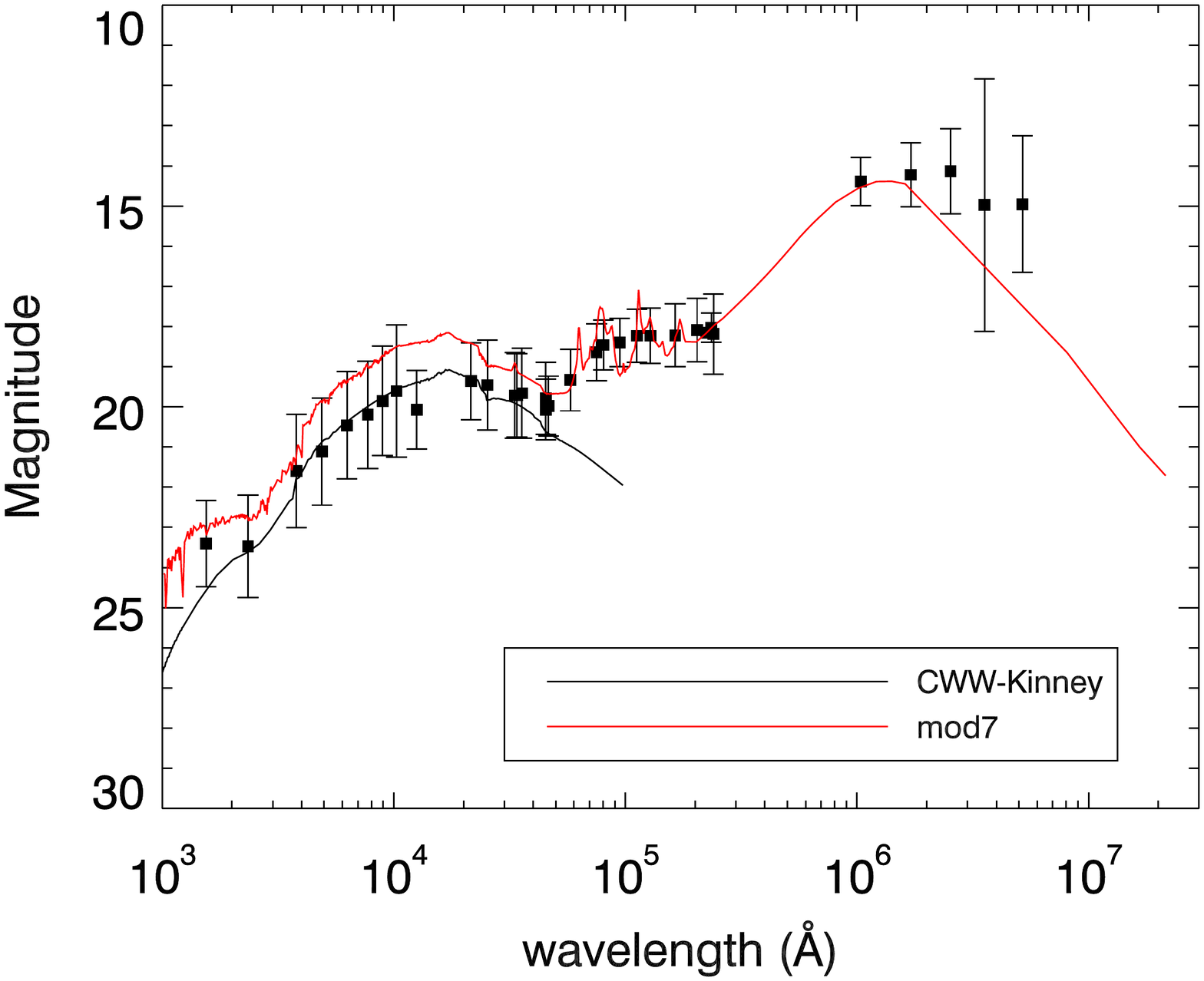}
	\includegraphics[width=\columnwidth]{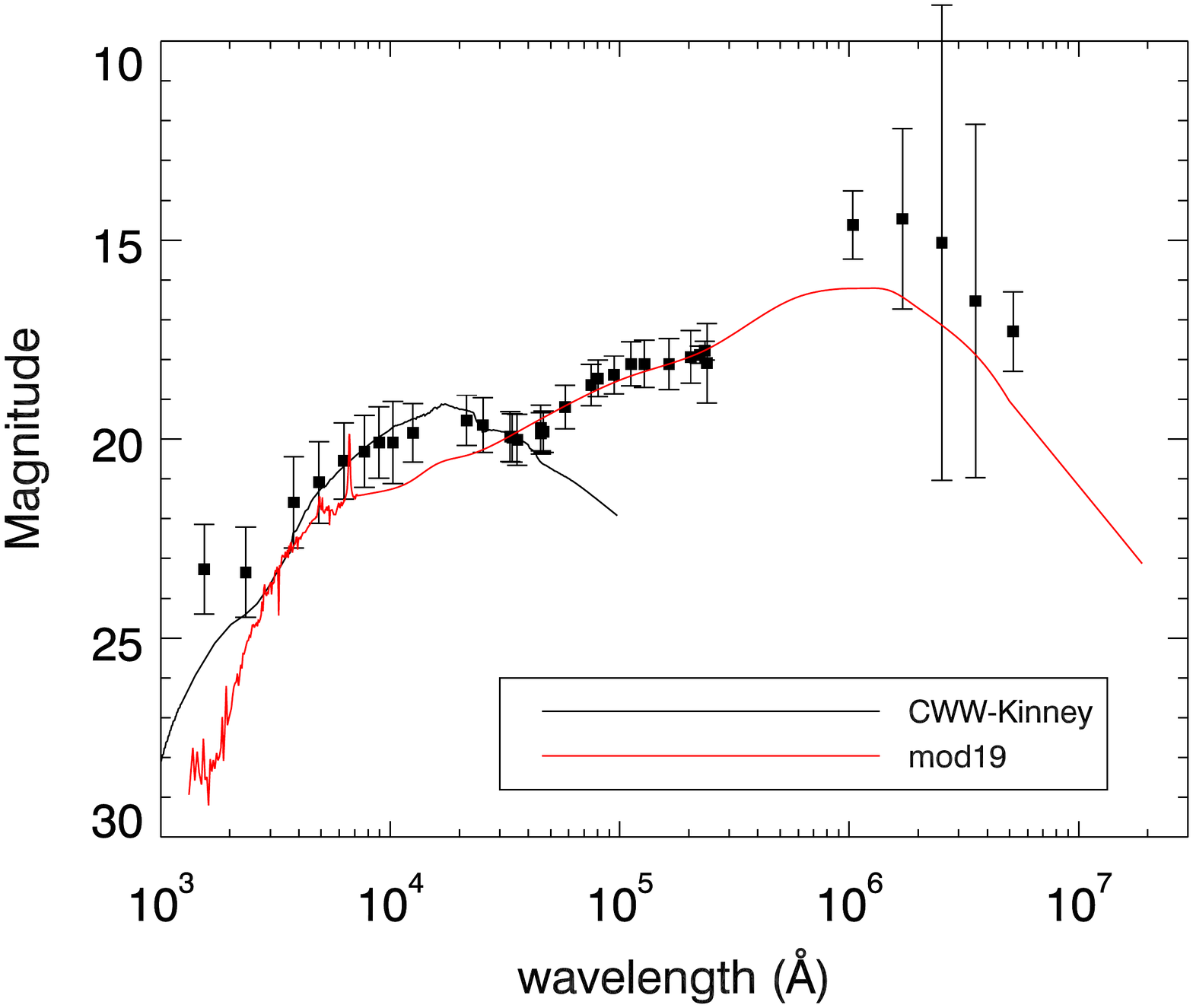}  
    \caption{Results of fitting median magnitudes of X-ray sample, SED SFG and SED AGN with CWW-Kinney (black) and SWIRE templates (red). The data points are plotted in black square with errors of one standard deviation. We do not show which specific model is fit in CWW-Kinney library in the figure.  \textsl{Top:} The SED fitting result of the median magnitude in X-ray sample, of which the best-fitting model is number 18 (Seyfert 2) in Table~\ref{tab:model}. \textsl{Middle:} The SED fitting of SED SFG median magnitude, of which the best-fitting model is number 7 (Spiral c) in Table~\ref{tab:model}. \textsl{Bottom:} the SED fitting of SED AGN median magnitude, of which the best-fitting model is number 19 (Type-2 QSO) in Table~\ref{tab:model}. }
   \label{fig:simu}
\end{figure}

To test the examination from the X-ray catalogue is valid or not, we stacked the magnitude of every band for all the X-ray objects and examined how they distribute statistically. We transfered all the observed magnitudes of objects from redshift 0.2 to 0.8 into the rest frame and then normalized them with 18$\mu$m magnitude. When the median magnitudes were calculated, the bin size of wavelength was chosen to be the same as the range of the bands used in our 18-band photometry. With the errors of the median magnitudes which are simply estimated by the standard deviation, we fit the median magnitudes again with the SED templates and see how they are classified. We assume the redshift to be 0.01 and the normalized magnitude zero point to be 18.5 in the SED fitting. The result is shown in the top panel of Fig~\ref{fig:simu}. The SED best-fitting model is 18, which is a "Seyfect 2" model in Table~\ref{tab:model}. As a result, to examine the IR AGN selection by matching X-ray catalogue is valid. 

\begin{figure}
     	\includegraphics[width=\columnwidth]{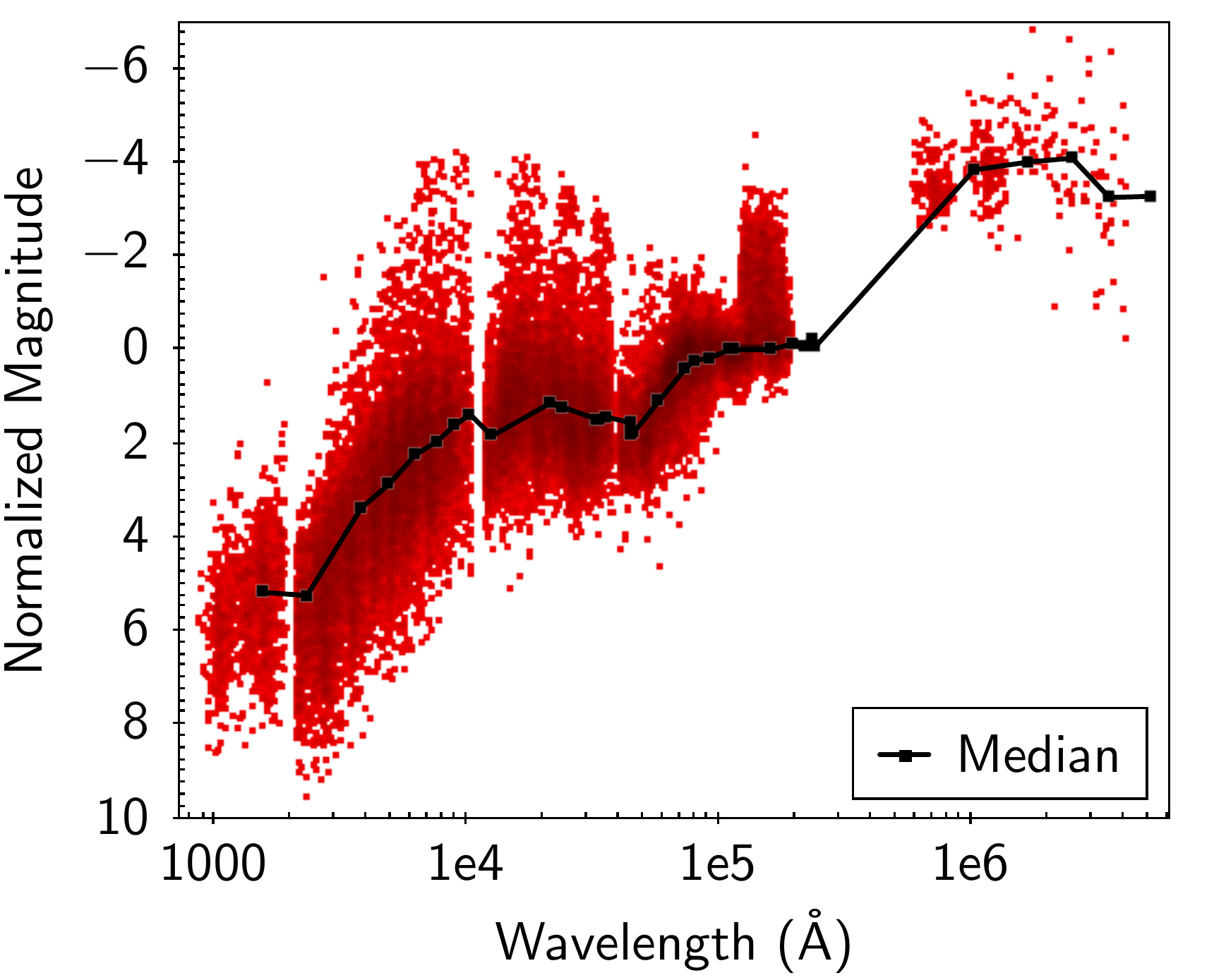}                 
        \includegraphics[width=\columnwidth]{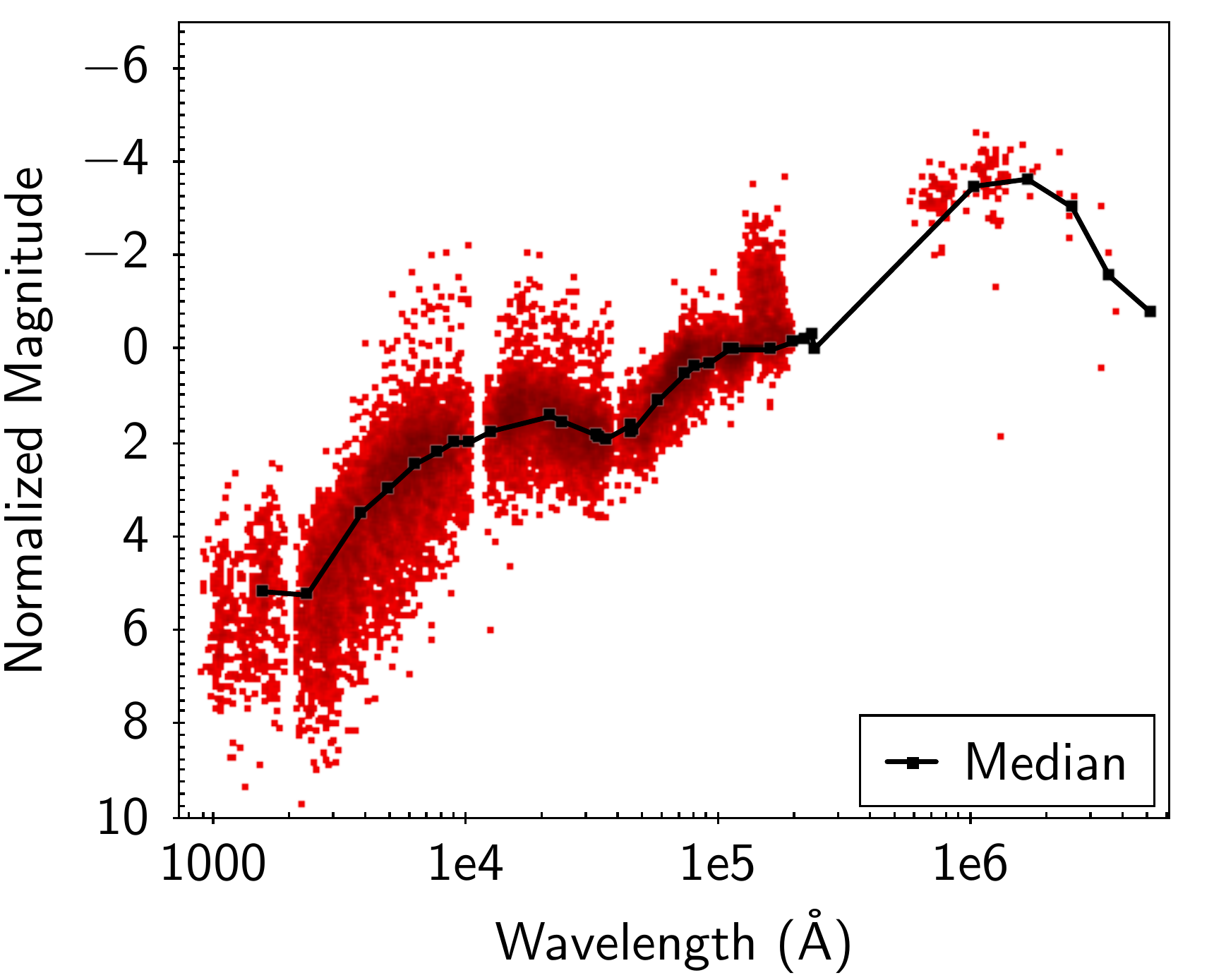}          
        \includegraphics[width=\columnwidth]{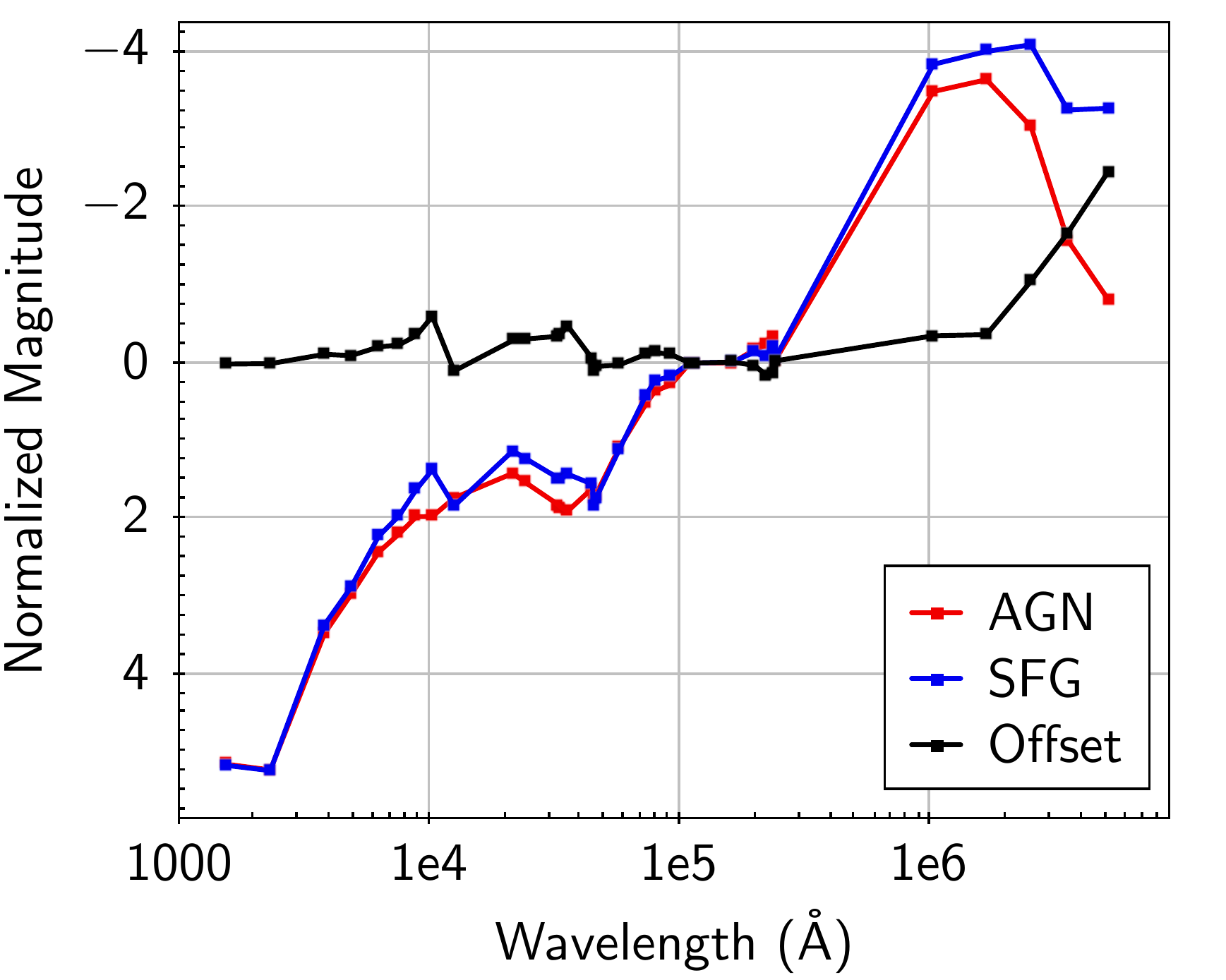} 
    \caption{The stacked median magnitudes are shown in this figure. $Top$: The stacked normalized magnitudes of SED SFG in red points, and the median in each band is plotted in black. The magnitudes are normalised by AKARI's 18$\mu$m bands. $Middle$: The same plot as the top panel for SED AGN. $Bottom$: The medians of SED SFG and SED AGN together. The SED AGN and SED SFG are plotted in red and blue, separately, and the offset between them is plotted in black.}
   \label{fig:SEDs}   
\end{figure}

In addition, we also performed the stacking analysis above for SED SFG and SED AGN sample, to check the consistency. The stacked magnitudes of SED SFG and SED AGN and their medians are shown in Fig.~\ref{fig:SEDs}. Stacked magnitudes of SFG and AGN are plotted in the top and middle panel respectively. The bottom panel shows the median magnitudes of SFG and AGN independently. There is a clear difference between those two median SEDs. The SFG median SED (blue) is brighter in both far infrared and near infrared, due to its star formation activity and stellar light. In mid-infrared, the strength of AGN's warm dust radiation and SFG's PAH emissions look comparable, but this could have a bias since we normalized magnitudes at 18 $\mu$m. Like the previous analysis on X-ray AGN, we fit these two SEDs with the templates, and set redshift of 0.01 and zero point magnitude 18.11, 18.23 for AGN and SED, respectively. The results are shown in the middle and bottom panels of Fig~\ref{fig:simu}. The FIR magnitudes have large errors due to its small sample size. The SFG median magnitudes fit to the model 7, which is the type "spiral c", while the AGN median magnitudes fit to the model 19, which is the "Type-2 QSO". It suggests that the AGN selection by our SED fitting is credible, because the AGN and SFG we selected have clear difference in median SED and the median SED are classified back to where it should be after fitting the template.

\section{Discussion}
\label{discussion}

In this paper, we present a new AGN selection method using the 18-band SED fitting in mid-IR. We selected more AGN than previous colour-colour selection by $\sim20\%$ . However, still many improvements can be done in the future and uncertainties need to be discussed.

\subsection{SED AGN missed in colour method}
\begin{figure}
	\includegraphics[width=\columnwidth]{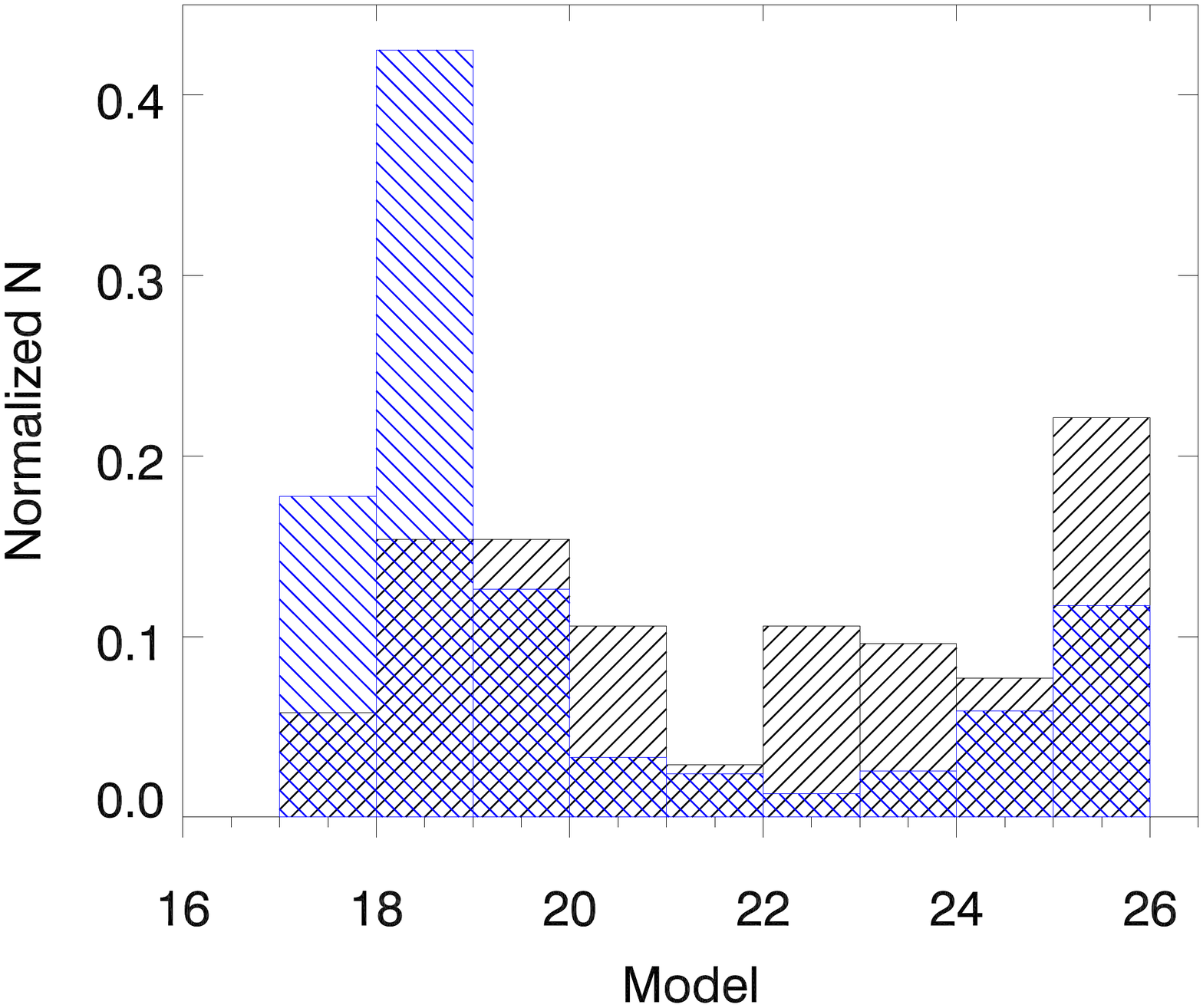}
    \caption{The normalized distribution of the best-fitting models of the SED AGN sample. The colour SFG distribution is filled with blue lines, while the colour AGN distribution is filled with black lines. See Table~\ref{tab:model} for the model numbers.}
    \label{fig:color_SFG_AGN_mod}
\end{figure}

\begin{figure}
	\includegraphics[width=\columnwidth]{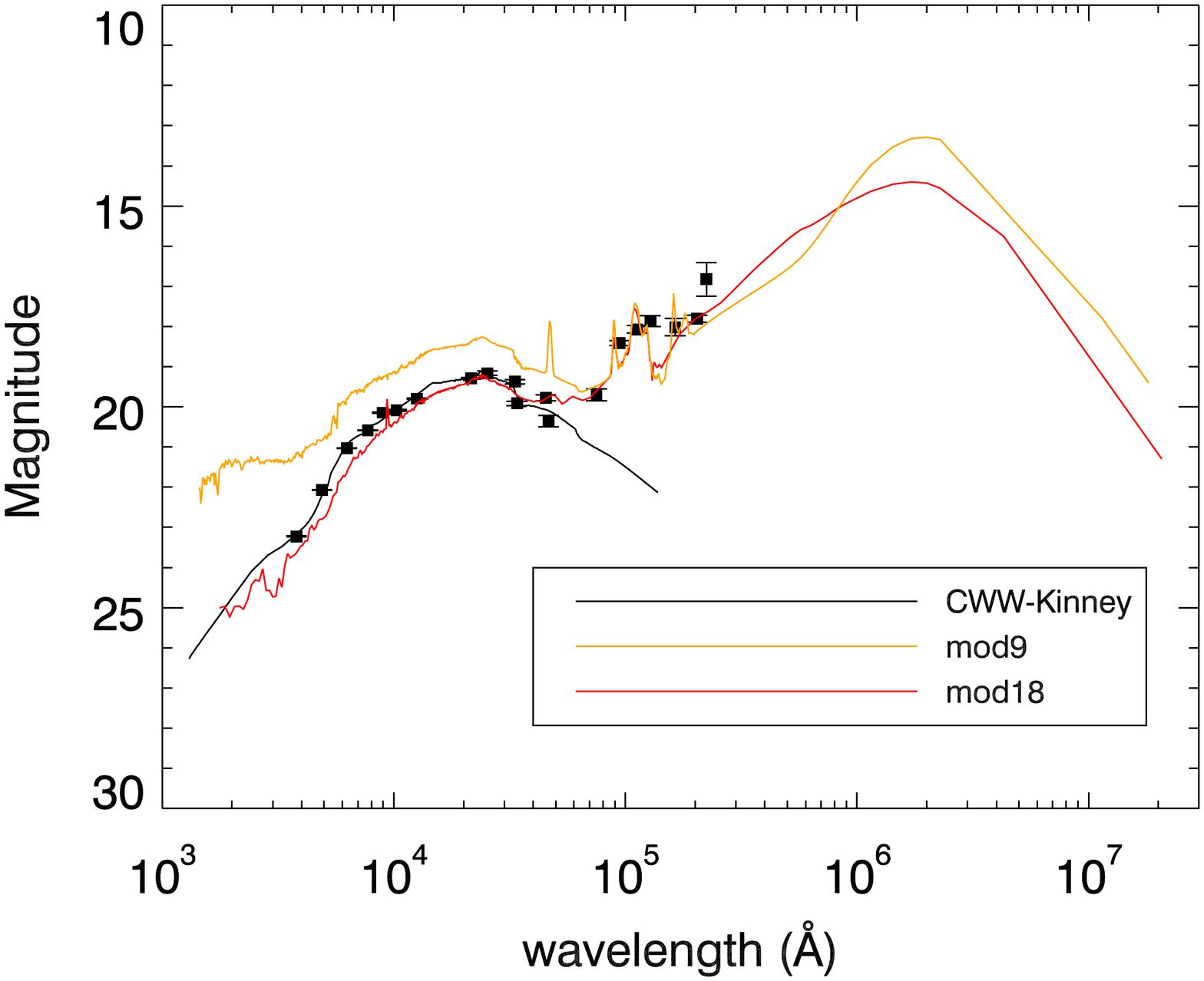}
    \caption{The SED of Id4869. The black curve is the best-fitting stellar component in the CWW-Kinney model. The best-fitting model in SWIRE template is model 18 (red), which is type "Seyfert 2". The best-fitting model in the SFG-only template is model 9 (orange), which is type "Spiral dm" in Table~\ref{tab:model}.}
    \label{fig:Id4869}
\end{figure}

In Fig.~\ref{fig:ccdiagram}, one can see that there are many SED AGNs outside the colour box, and most of them are fit with the model 17 and 18. We discuss those SED AGNs missed in colour method here again. In this section, we only focus on SED AGN and compare the model distributions of the two classifications by colour (colour SFG and colour AGN). The normalized best-fitting model distributions of colour AGN and colour SFG in SED AGN are shown in Fig.~\ref{fig:color_SFG_AGN_mod}. The SED AGNs missed by colour box are mostly model 17 and 18. These two models are "Seyfert 1.8" and "Seyfert 2". Seyfert models in SWIRE template are created by moderately luminous or host-dominated AGNs and they have redder WISE $[3.4]-[4.6]$ colour than the criteria. That is why the colour method cannot capture them. We provide the SED of Id4869 as an example in Fig.~\ref{fig:Id4869}. It is classified into model 18 which is Seyfert 2 in SED fitting with $\chi^{2}=28.64$, while it locates outside of the WISE box with colours $[3.4]-[4.6]=0.2$ and $[4.6]-[12]=4.3$. We performed the SED fitting again with only SFG templates (i.e. model 1 to 17) so that by comparing with the best-fitting models, we can see if the SED fitting really separate Seyfert from SFG or not. The Id4869 has redshift 0.43, so there are 9 bands used in MIR fitting. The best-fitting SFG template is model 9 with $\chi^{2}=41.29$. The data fit with the model 18 (red) better than model 9 (orange) at 4.6$\mu$m, 7$\mu$m, and 9$\mu$m, where the PAH feature of model 9 is deviating from the data points. This shows an example that the colour method misses host-dominated AGNs, while SED fitting can capture them successfully.

\subsection{Colour AGN missed in SED fitting}
\begin{figure*}
	\includegraphics[width=\columnwidth]{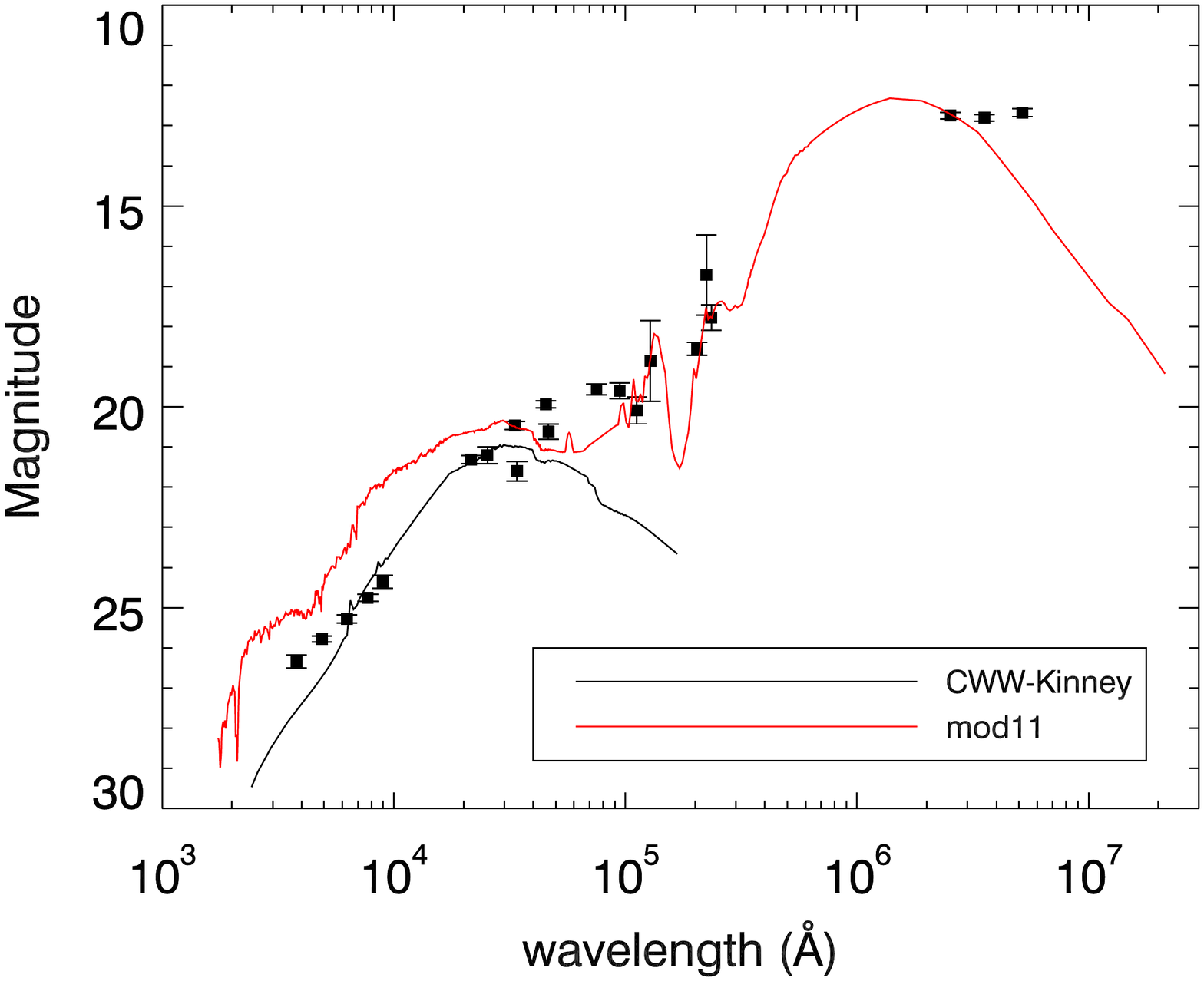}
	\includegraphics[width=\columnwidth]{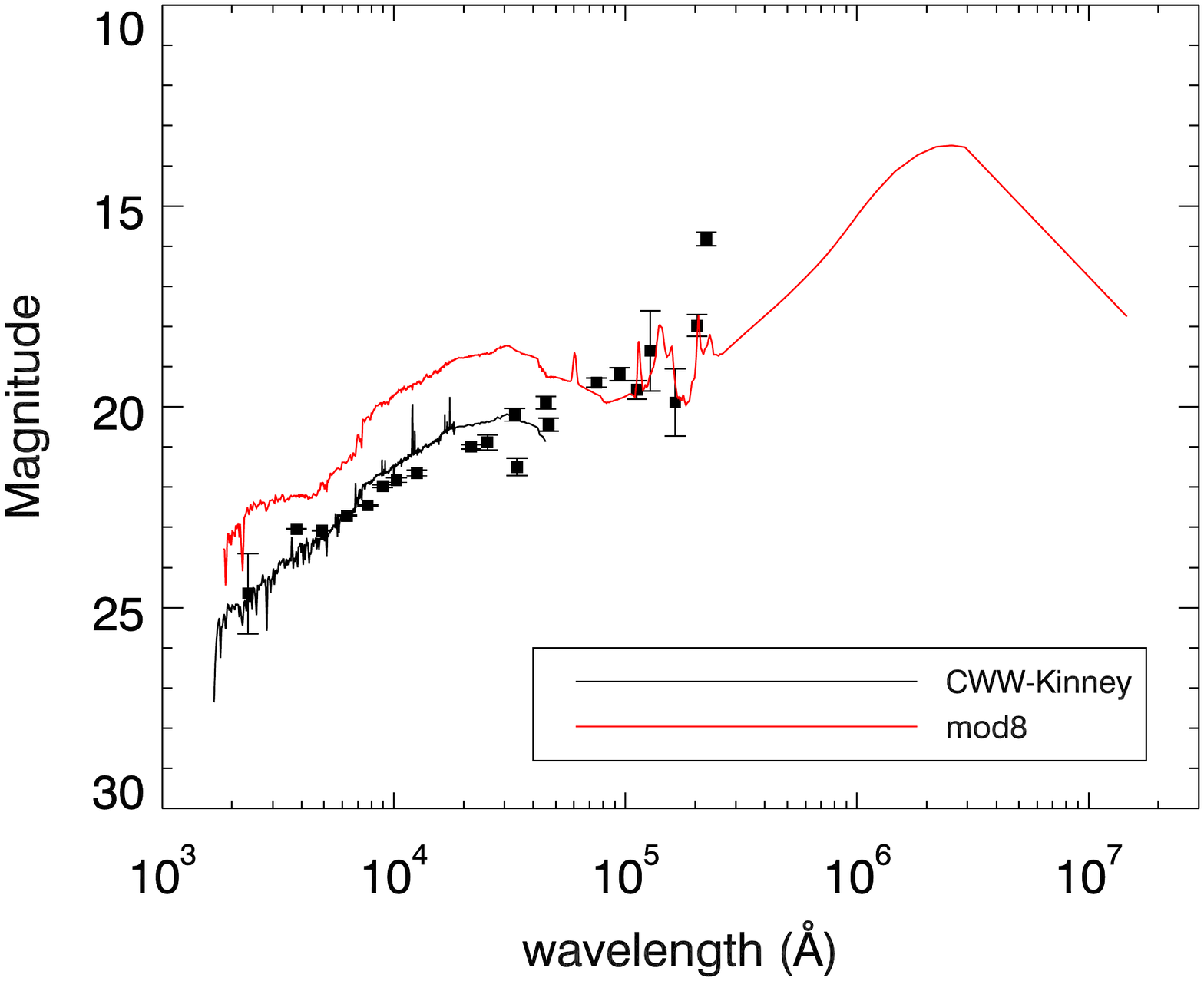}
	\includegraphics[width=\columnwidth]{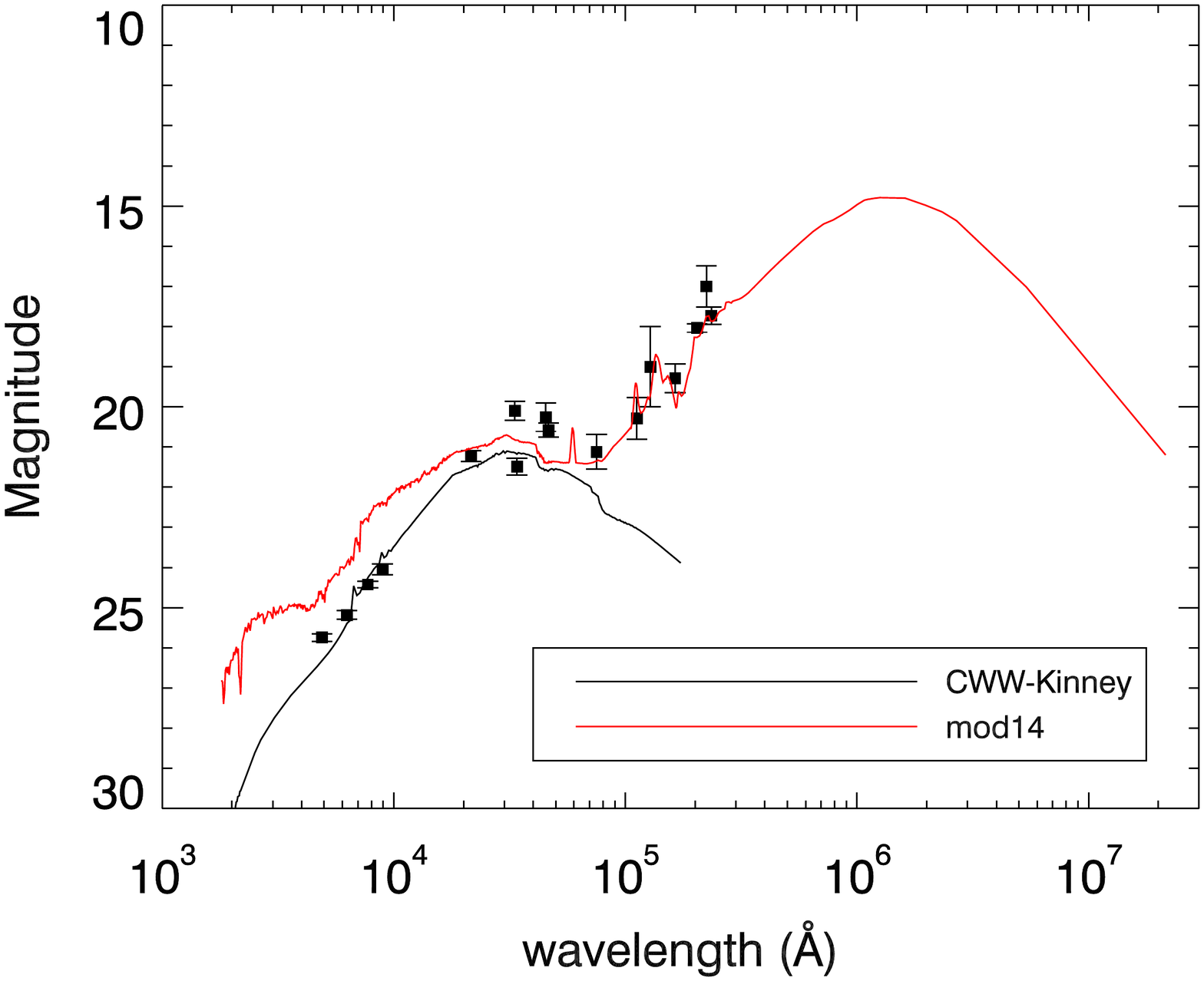}
	\includegraphics[width=\columnwidth]{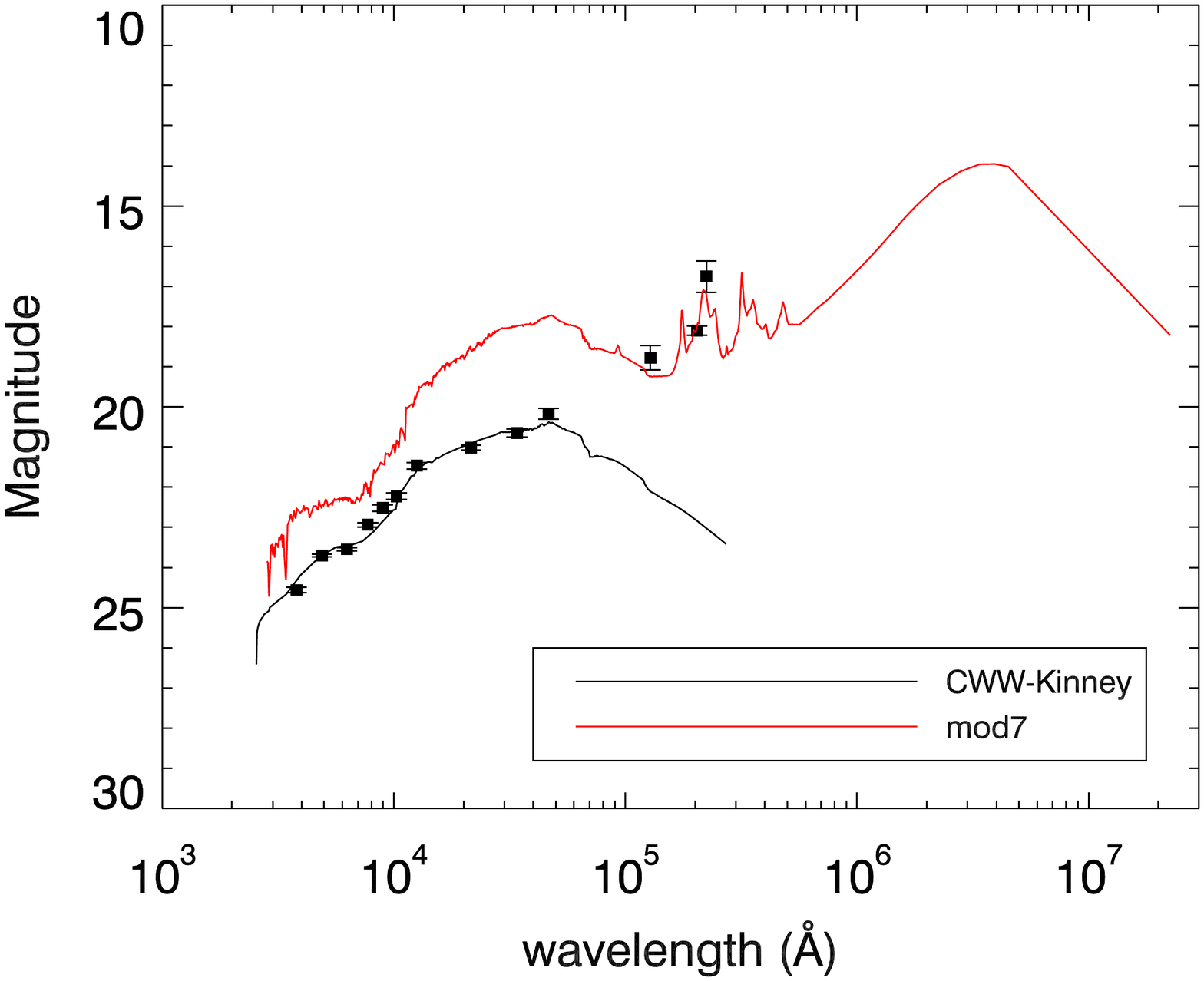}
	\includegraphics[width=\columnwidth]{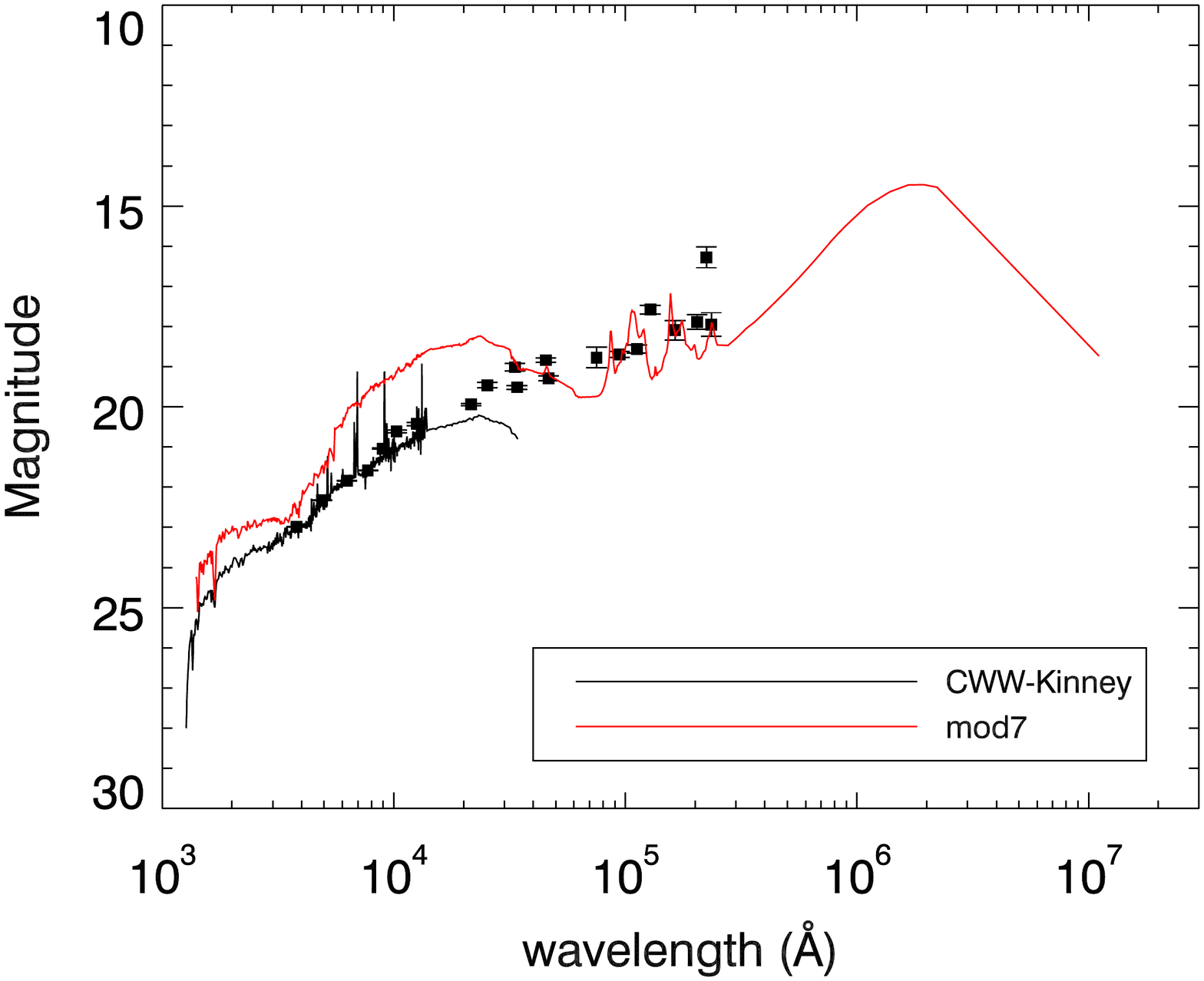}
	\includegraphics[width=\columnwidth]{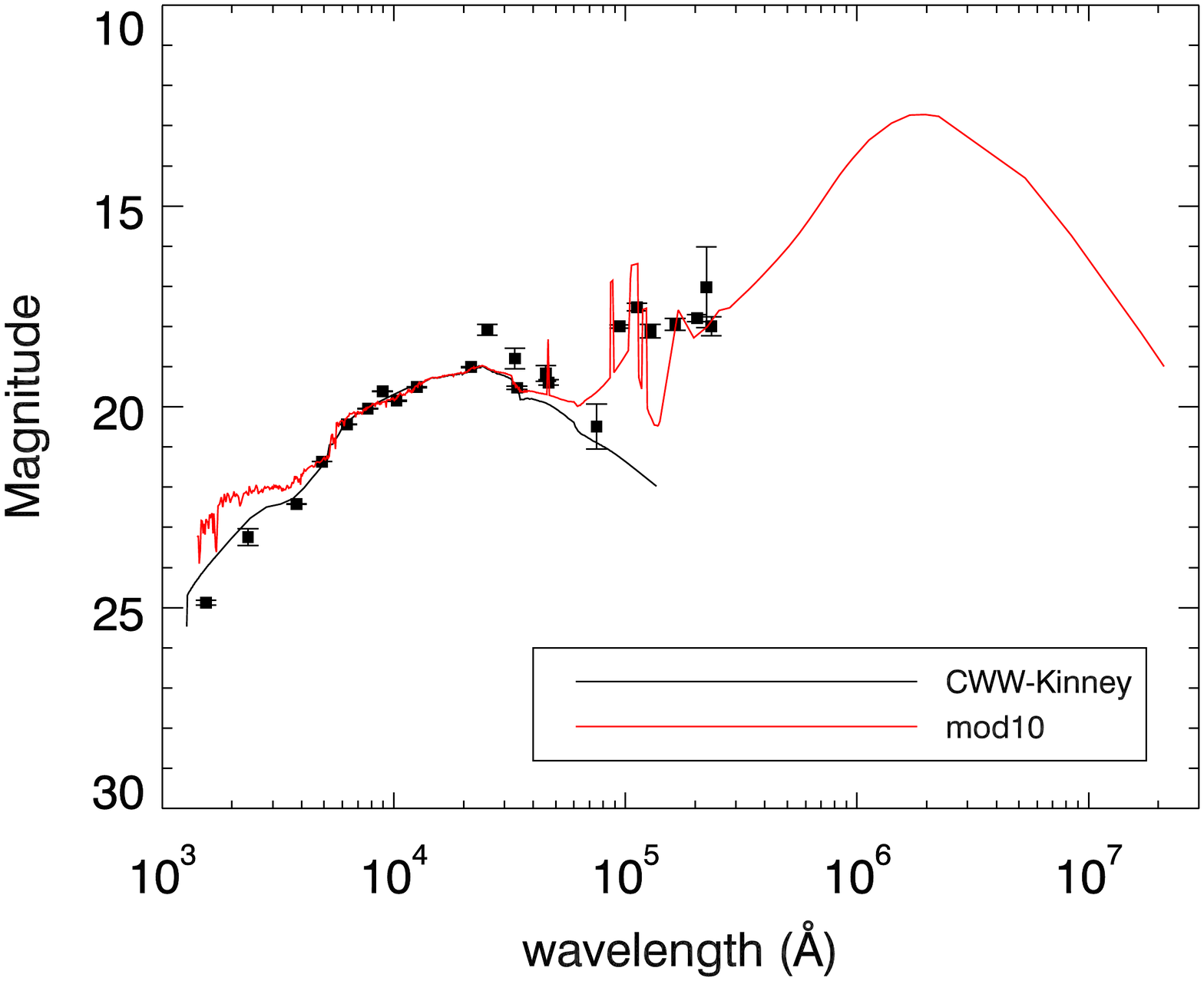}
    \caption{Example SEDs of colour AGNs that are fit to be SFG in SED fitting.}
    \label{fig:colourAGN}
\end{figure*}

On the contrary, there are many SED SFGs inside the WISE colour box. This inconsistency between two IR selections is perplexing to us. Not having a definite answer to the problem, we can just propose some possible conjectures. First, we provide some SEDs of those colour AGN in Fig.~\ref{fig:colourAGN} to convince that even though the results are not consistent, our fitting is still valid. These objects do not have obvious AGN power law feature but have PAH emission feature peaks, so they are more likely to be SFG. In this sense, selection by 2 colours could have some failures because of their mere 3 bands. However, our SED fitting selection may have a weakness at high redshift. By examing the catalogue and SEDs, we found that the colour AGNs generally have higher redshift than SED AGNs in both average and median (e.g. average z$\sim$1.05 for colour AGNs and average z$\sim$0.69 for SED AGNs). At z$\sim$1, the  WISE [3.4] and [4.6] bands observe 1.7 and 2.3 $\mu$m in rest frame. These bands are not used in the IR fitting for AGN selection in our method because they have been fit as the stellar component and have been substracted. Also, because of the shifting of wavelengths at high redshift, the high redshift objects are observed at shorter wavelengths. This means that we lose our template's advantage to capture mid-IR light at $\sim10~\mu$m. The conclusion here is that the colour method may have some failures in the AGN selection becuase of its limited number of bands, but on the other hand our SED fitting may have poorer performance at high redshift as well.      

\subsection{Implication for Luminosity Function of AGN}
According to the Sect.~\ref{AGN} and Table~\ref{tab:18-band},  there are 650 SED AGNs and 287 colour AGNs in the colour sample. The number of SED AGN is roughly 2 times more than colour AGN, which implies that the previous studies possibly have under-estimated the number of AGN in the universe. The difference mainly comes from the Seyfert models (model 17 and 18), which implies the existence of those moderately luminous or host-dominated AGNs. Consequently, AGN luminosity function and black hole accretion rate density may be higher than the calculation by previous work. Therefore, this implication is important in cosmology and galaxy evolution.  However, further investigations such as luminosity function are still needed to conclude more accurately.   

\subsection{Relaxed selection and AGN fraction}
\begin{figure}
	\includegraphics[width=\columnwidth]{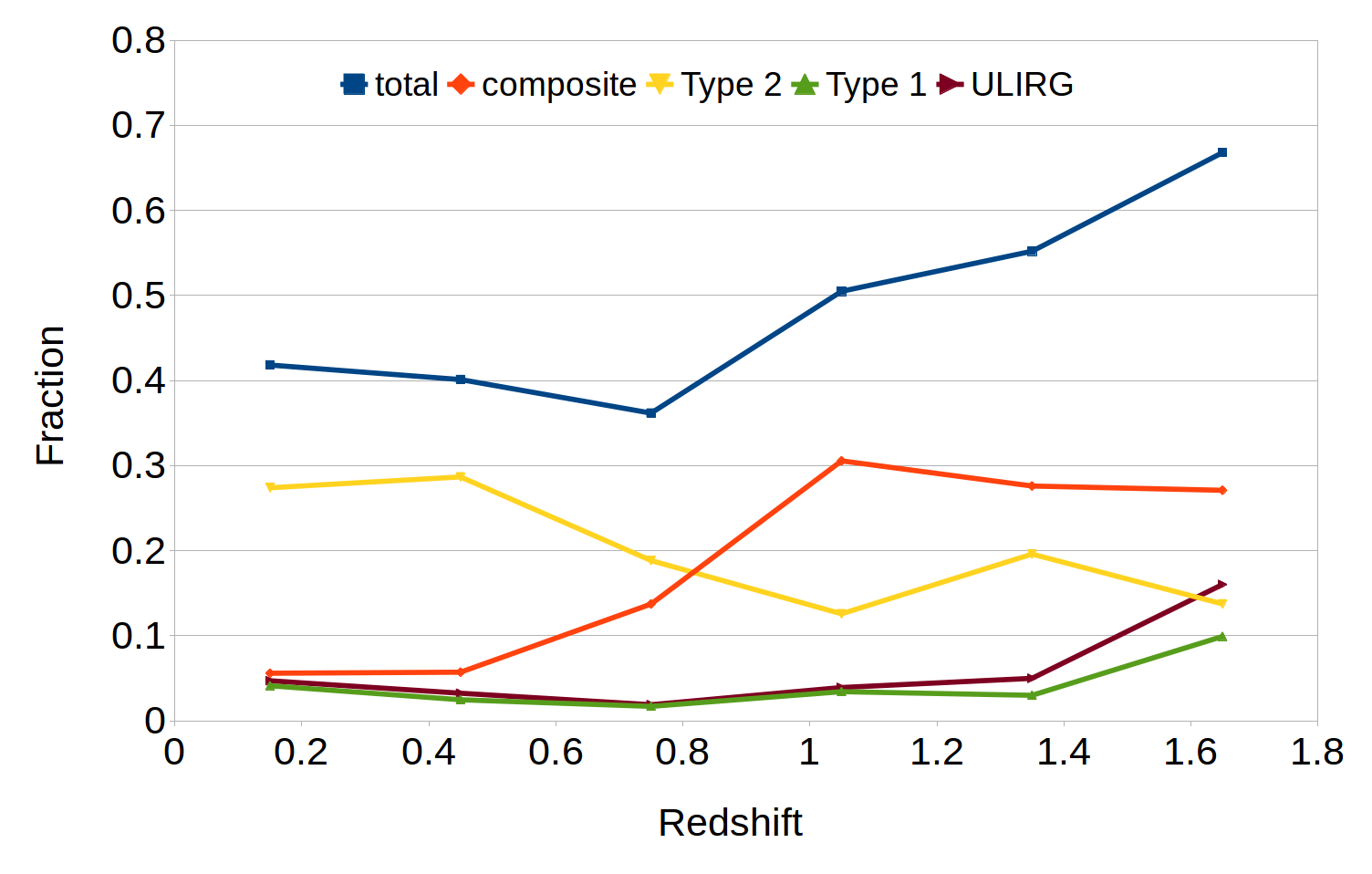}
    \caption{AGN fractions as functions of redshift. The orange line is composites, which defined by model 11, 12, 13 and 15. The yellow line is Types 2 (model 17, 18, 19, and 25). The green line is Types 1 (model 20, 21, and 22). The purple line is ULIRG (model 23 and 24). The blue line contains all the components above.}
    \label{fig:AGN_fraction}
\end{figure}

From our result in Sect.~\ref{AGN}, we found that the AGN fraction in the AKARI NEP deep field is $29.6\pm0.8\%$. Nevertheless, the result of AGN selection depends on its definition. We discuss that it is an option to use more relaxed definition, although it may enhance the uncertainties from AGN/SFG composites.  \citet{Gruppioni et al. 2013} defines a relaxed AGN selection for SWIRE template. They regard our model 12, 13, and 15, which are in the Starburst/ULIRG or Starburst/Seyfert2 category, as the SFG containing obscured AGN. Refering to their study, we used this definition and also included the model 11 as a relaxed AGN selection because it is in the Starburst/ULIRG category as well. In this discussion, we define our model 11, 12, 13, and 15 to be starburst/ULIRG composite. Model 17, 18, 19 and 25 are Type 2 and Model 20, 21, and 22 are Type 1. The ULIRG is defined by model 23 and 24. If the starburst/ULIRG templates are counted as AGN, the recovering rate becomes 46$\pm$5.5$\%$ at maximum redshift 0.9, which are 6$\%$ larger than our regular selection and 17$\%$ larger than the conventional colour-colour method. If we include the starburst/ULIRG composite templates in our selection, then the AGN fraction of our whole sample will become $44\pm1.0\%$. This number is consistent with previous work, for example, \citet{Feltre et al. 2013} and \citet{Kirkpatrick et al. 2015}. The former used Spitzer IRS data and found 45$\%$ mid-IR AGN by PAH equivalent width in HerMES field. The latter performed SED fitting with Spitzer and Herschel photometry and Spitzer IRS spectroscopy. They selected IR AGN by fitting IR SED into spectral decompositions and present 30$\%$ SFGs, 34$\%$ composites, and 36$\%$ AGNs. They suggest $>40\%$ of IR populations have AGN activity. We show the AGN fraction as functions of redshift in Fig.~\ref{fig:AGN_fraction} with four classifications, composite, Type 2, Type 1, and ULIRG. In the AKARI NEP deep field, AGN fraction is 0.42$\pm$0.04 at low redshift, and increases to 0.67$\pm$0.06 at redshift 1.5, which suggests that AGN activity is stronger in distant universe or at larger IR luminosity. 

\subsection{Comparision with another template}
\begin{figure}
	\includegraphics[width=\columnwidth]{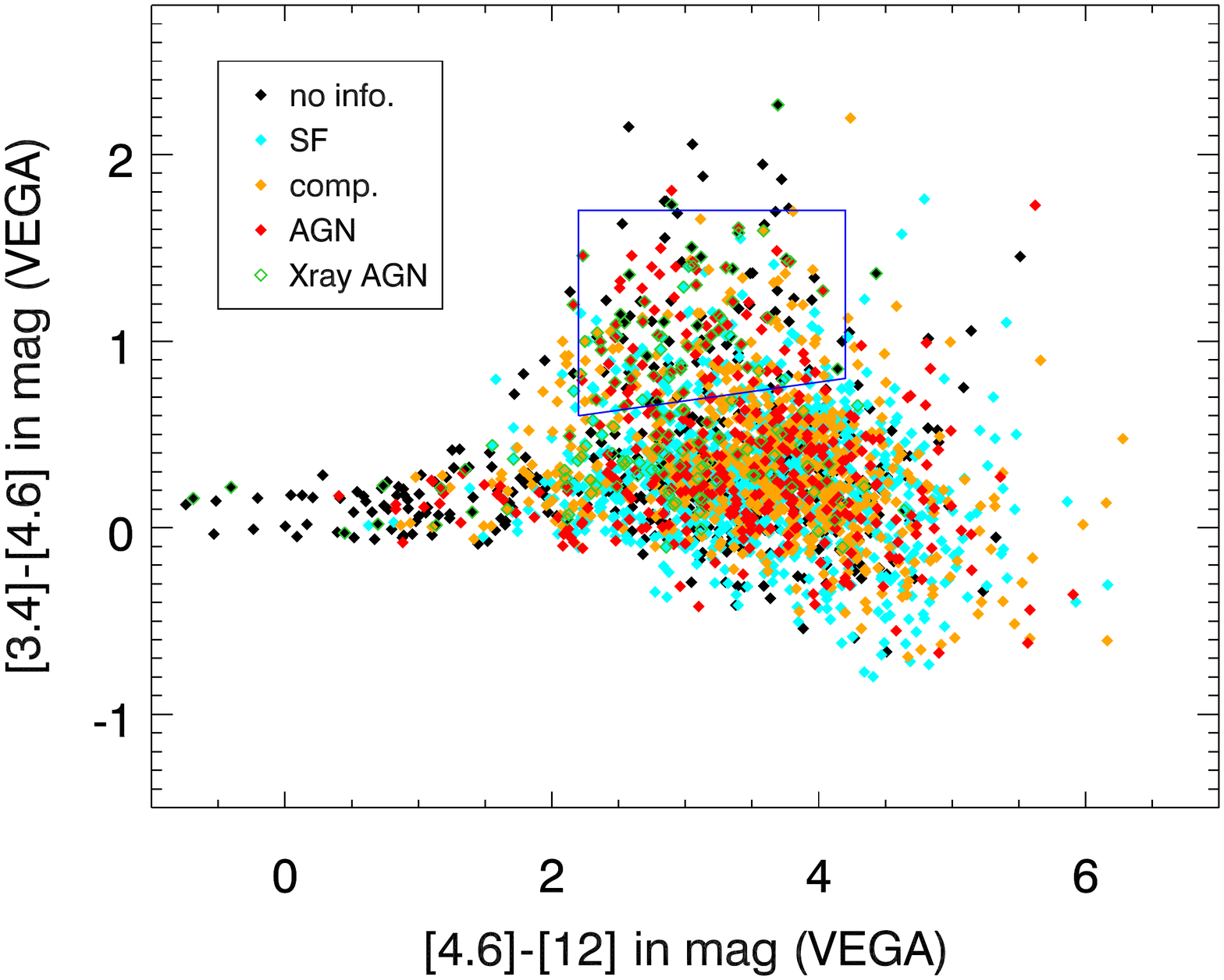}	
	
    \caption{The WISE colour-colour diagram of the colour sample with the AGN classification with the template from \citet{Brown et al. 2014}. AGNs are plotted in red, while SFGs are plotted in cyan. There are also AGN/SFG composites plotted in orange. The objects without classification are plotted in black.}
    \label{fig:ccdiagram_BROWN}
\end{figure}

\citet{Brown et al. 2014} provides a set of SEDs with 129 nearby galaxies. They classify those 129 galaxies into 50 SFGs, 16 AGNs, and 31 composites by BPT diagram. The remaining 32 galaxies have no classification. Even though the Polletta's template and Brown's template may have some intrinsic difference, we performed the SED fitting with Brown's template to check the consistency of our method and results. We used the same criteria to create the SED sample as we did in Section.~\ref{SED}. In the 4833 objects in the SED sample using Brown's template, 770 are classifed into AGN, 1191 are composites, 1984 are SF, and 888 have no information. The AGN fractions are 15.9$\pm$0.6$\%$ and 40.6$\pm$0.9$\%$ if we include composites. From this comparison, our previous results (29.6$\pm$0.8$\%$) with Polletta's template looks reasonable since we may have selected some composites from the Seyfert models. Fig.~\ref{fig:ccdiagram_BROWN} shows the WISE colour-colour diagram. Similar to our previous result (Fig.~\ref{fig:ccdiagram}), this shows that AGN and composites distribute widely from the AGN criteria box of WISE colours.

\subsection{Future prospects}
Most results in this work are based on the comparison of the X-ray AGN recovering rate. However, the X-ray sample contains only 457 objects, which is still limited. Hence, to have more X-ray survey in NEP field, preferably in hard X-ray, would be very helpful. Also, we are looking forward to having new IR telescopes like JWST and SPICA, which have better visibility, to perform observations in the NEP field, too. 

According to our investigation on the number of bands (Fig.~\ref{fig:recovering_rate}), 9 bands apparently has higher recovering rate than 3 bands, which shows that having more bands is  undoubtedly better. We suggest that if any research is going to perform SED fitting, then having at least 9 bands will be more favourable. Having a new project or new telescope is in the same situation. It is beneficial to use or build a telescope equipped with at least 9 filters.    

This work is the first step of AGN researches in the AKARI NEP deep field. We plan to perform further research like evolution of AGN luminosity funtion and/or accretion rate by this method in future. Additionally, as more and more surveys cover the NEP field, such as Subaru HSC \citep{Goto et al. 2017}, we are going to accumulate more knowledge in distant universe and understand how the universe evolves.

\section{Summary}
\label{summary}
AKARI IRC 9 filters provide us with continuous 9 bands in mid-IR, which only AKARI can provide. Using the AKARI NEP deep field data and its multi-wavelength counterpart from GALEX, CFHT, WISE, Spitzer and Herschel, we selected AGN by SED fitting. Based on the mid-IR features of AGN and SFG, AGN is selected by the best-fitting model in SWIRE template, which includes 16 galaxy and starburst models and 9 AGN models. Matching with the Chandra Xray catalogue, we concluded that our selection by SED fitting recovers more AGNs than previous colour criteria by $\sim$20 percentage points. The reliability of the AGN selection by SED fitting has been tested by stacking all the normalized rest-frame magnitudes of AGN and SFG. These two medians of stacked magnitudes show the significant difference, and suggesting our methodology works. Our 18-band AGN selection recovered a factor of 2 more AGN than previous colour-colour selection, possibly altering the cosmic AGN accretion history.

\section*{Acknowledgements}
We thank the anonymous referee for reviewing thoroughly and providing many constructive comments for this paper. We thank our group member Yi-Han Wu and collaborator Ji-Jia Tang for checking and giving many useful advices. This research is based on observations with AKARI, a JAXA project with the participation of ESA. Also, we appreciate the observations from WISE, Spitzer, GALEX, CFHT, Herschel, and Chandra for our cross-matching. This work was supported by the grant 105-2122-M-007-003-MY3 from Ministry of Science and Technology of Taiwan (MOST).

%%%%%%%%%%%%%%%%%%%%%%%%%%%%%%%%%%%%%%%%%%%%%%%%%%



%\bibliographystyle{mnras}


%%%%%%%%%%%%%%%%%%%%%%%%%%%%%%%%%%%%%%%%%%%%%%%%%%

%%%%%%%%%%%%%%%%% APPENDICES %%%%%%%%%%%%%%%%%%%%%

\appendix

\section{SWIRE template}
The SWIRE template library is an empirically based one. It contains 3 ellipticals, 7 sprirals, 6 Starbursts (4 of them are starburst/ULIRG composites), 7 AGNs, and 2 AGN-Starburst composite. Templates range from 1000 {\AA} to 1000~$\mu$m. Elliptical, spiral and starburst templates were generated by the GRASIL code. Seyfert 1.8 and Seyfert 2 templates describe AGNs with moderate luminosity. They were created by combining models with observational photometry from NED and spectroscopy from ISO. QSO1 templates were derived by optical quasar spectrum (SDSS) and IR data from SDSS/SWIRE quasar sample. One QSO2 template was obtained by the optical and near-IR spectrum of the red quasar FIRST J013435.7$-$093102 and the IR data from Palomar-Green sample. Another QSO2 is torus, which is the SED of a type 2 QSO SWIRE$\_$J104409.95$+$585224.8. There are three composite templates which are the best-fitting SED to QSO Mrk 231, Seyfert 2 galaxies IRAS 19254$-$7245 South, and IRAS22491$-$1808. More detail can be found in \citet{Polletta et al. 2007} and the SWIRE template library website\footnote{\url{http://www.iasf-milano.inaf.it/~polletta/templates/swire_templates.html}}.

%%%%%%%%%%%%%%%%%%%%%%%%%%%%%%%%%%%%%%%%%%%%%%%%%%

% Don't change these lines
\bsp	% typesetting comment
\label{lastpage}
\end{document}